\documentclass[reprint,aps,graphicx,pre,showpacs]{revtex4-1}
\usepackage{amsmath,amssymb,amsfonts}   
\usepackage{graphicx,bm}
\usepackage{paralist}
\usepackage{epstopdf}
\usepackage{graphics} 
\usepackage[colorlinks=true]{hyperref}
\hypersetup{urlcolor=blue, citecolor=red}
\usepackage[latin1]{inputenc}

\renewcommand{\theequation}{\arabic{section}.\arabic{equation}}

\newcommand{\calU}{{\mathcal U}}
\newcommand{\calG}{{\mathcal G}}
\newcommand{\R}{{\mathbb R}}

\newcommand{\X}{\mathbf{X}}
\renewcommand{\P}{\mathbb{P}}
\newcommand{\PP}{\widetilde{P}}
\newcommand{\x}{\mathbf{x}}
\newcommand{\y}{\mathbf{y}}
\newcommand{\e}{{\mathrm e}}
\newcommand{\E}{{\mathbb E}}

\newcommand{\n}{\mathbf n}
\newcommand{\calT}{{\mathcal T}}
\renewcommand{\P}{\mathbb P}
\newcommand{\p}{\widetilde{p}}
\renewcommand{\v}{\widetilde{v}}
\renewcommand{\u}{\widetilde{u}}
\newcommand{\U}{\widetilde{U}}
\newcommand{\V}{\widetilde{V}}
\newcommand{\ellh}{\hat{\ell}}

\begin{document}

 \title{The narrow capture problem: an encounter-based approach to partially reactive targets}

%\author{ \em P. C. Bressloff, \\ Department of Mathematics, University of Utah \\155 South 1400 East, Salt Lake City, UT 84112}

\author{Paul C. Bressloff}
\address{Department of Mathematics, University of Utah 155 South 1400 East, Salt Lake City, UT 84112}

\begin{abstract} 
A general topic of current interest is the analysis of diffusion problems in singularly perturbed domains with small interior targets or traps (the narrow capture problem). One major application is to intracellular diffusion, where the targets typically represent some form of reactive biochemical substrate. Most studies of the narrow capture problem treat the target boundaries as totally absorbing (Dirichlet), that is, the chemical reaction occurs immediately on first encounter between particle and target surface. In this paper, we analyze the three-dimensional narrow capture problem in the more realistic case of partially reactive target boundaries. We begin by considering classical Robin boundary conditions. Matching inner and outer solutions of the single-particle probability density, we derive an asymptotic expansion of the Laplace transformed flux into each reactive surface in powers of $\epsilon$, where $\epsilon \rho$ is a given target size. In turn, the fluxes determine the splitting probabilities for target absorption. We then extend our analysis to more general types of reactive targets by combining matched asymptotic analysis with an encounter-based formulation of diffusion-mediated surface reactions.
That is, we derive an asymptotic expansion of the joint probability density for particle position and the so-called boundary local time, which characterizes the amount of time that a Brownian particle spends in the neighborhood of a point on a totally reflecting boundary. The effects of surface reactions are then incorporated via an appropriate stopping condition for the boundary local time. Robin boundary conditions are recovered
 in the special case of an exponential law for the stopping local times. Finally, we illustrate the theory by exploring how the leading-order contributions to the splitting probabilities depend on the choice of surface reactions. In particular, we show that there is an effective renormalization of the target radius of the form $\rho\rightarrow \rho-\widetilde{\Psi}(1/\rho)$, where $\widetilde{\Psi}$ is the Laplace transform of the stopping local time distribution.

\end{abstract}
%%%%%%%%%%%%%%%%%%%%%%%%%%%

\maketitle
\section{Introduction}

A topic of increasing interest is the analysis of two-dimensional (2D) and three-dimensional (3D) diffusion in singularly perturbed domains \cite{Ward93,Ward93a,Ward00,Straube07,Schuss07,Bressloff08,Benichou08,Coombs09,Pillay10,Reingruber10,Cheviakov10,Cheviakov11,Chevalier11,Holcman14a,Ward15,Coombs15,Bressloff15,Bressloff15a,Lindsay15,Lindsay16,Lindsay17,Grebenkov17,Bressloff21a,Bressloff21b,Bressloff22a}. Two broad classes of problem are diffusion in a domain with small interior targets or traps, and diffusion in a domain with an exterior boundary that is reflecting almost everywhere, except for one or more small holes through which particles can escape. One major application of these studies is molecular diffusion within biological cells, where interior targets could represent (possibly reactive) intracellular compartments and holes on the boundary could represent ion channels or nuclear pores \cite{Holcman15,Bressloff22}. 
Quantities of interest at the level of bulk diffusion include the steady-state solution (assuming it exists) and the approach to steady state, as characterized by the leading non-zero eigenvalue $\lambda_1$ of the negative Laplacian \cite{Ward93,Ward93a,Ward00,Cheviakov11} or by the so-called accumulation time \cite{Bressloff22a}. In addition, the flux into an interior target can be used to determine an effective reaction rate \cite{Straube07,Bressloff15}. At the single-particle level, the solution of the diffusion equation (or more general Fokker-Planck equation) represents the probability density to find the particle at a particular location. One is now typically interested in calculating the splitting probabilities and conditional mean first passage times time for a particle to be captured by an interior target (narrow capture) \cite{Coombs09,Chevalier11,Ward15,Coombs15,Lindsay16,Lindsay17,Bressloff21a,Bressloff21b} or to escape from a domain through a small hole in the boundary (narrow escape) \cite{Schuss07,Benichou08,Pillay10,Cheviakov10,Reingruber10,Holcman14a,Bressloff15a,Lindsay15,Grebenkov17}. For all of these examples, the quantity of interest satisfies an associated boundary value problem (BVP), which can be solved using a mixture of matched asymptotic analysis and Green's function methods.

Within the context of narrow capture problems in cell biology, absorption by a target typically represents some form of chemical reaction. In almost all studies of diffusion in singularly perturbed domains, the boundary conditions imposed on the small targets are taken to be totally absorbing (Dirichlet). A totally absorbing target means that the only contribution to the effective reaction rate is the transport process itself, since the chemical reaction occurs immediately on first encounter between particle and target. In other words, the reaction is diffusion-limited rather than reaction-limited \cite{Rice85}. However, a more realistic scenario is to consider a combination of a transport step and a reaction step, both of which contribute to the effective reaction rate. Collins and Kimball \cite{Collins49} incorporated an imperfect reaction on a target surface $\partial \calU$ by replacing the Dirichlet boundary condition with the Robin or partially reflecting boundary condition
 \begin{equation*}
 -D\nabla c(\x,t)\cdot \n=\kappa_0 c(\x,t),\quad \x \in \partial \calU.
 \end{equation*}
 Here $c(\x,t)$ is particle concentration, $\n$ is the unit normal at the boundary that is directed towards the center of the target, $D$ is the diffusivity, and $\kappa_0$ (in units m/s) is known as the reactivity constant. The above boundary condition implies that there is a net flux of particles into the target (left-hand side), which is equal to the rate at which particles react with (are absorbed by) the target (right-hand side). The latter is taken to be proportional to the particle concentration at the target with $\kappa_0$ the constant of proportionality. The totally absorbing case is recovered in the limit $\kappa_0 \rightarrow \infty$, whereas the case of an inert (perfectly reflecting) target is obtained by setting $\kappa_0=0$. In practice, the diffusion-limited and reaction-limited cases correspond to the regimes $\xi \ll R$ and $\xi \gg R$, respectively. Here $R$ is a geometric length-scale such as the radius of a spherical target and $\xi=D/\kappa_0$ is known as the reaction length. Note that there have been a few studies of bulk diffusion in singularly perturbed domains containing targets with Robin boundary conditions \cite{Ward93,Ward93a,Ward00,Bressloff08}. It is also possible to obtain Robin boundary conditions by spatially homogenizing a target with mixed boundary conditions \cite{Lindsay15}, or by considering a stochastically-gated target in an appropriate limit \cite{Lawley15}. However, as far as we are aware, there have not been any detailed studies at the single-particle level.
 
 As recently highlighted by Grebenkov \cite{Grebenkov19a}, the single-particle probabilistic interpretation of the partially reflecting boundary condition is much more complicated than the Dirichlet boundary condition. The latter is easily incorporated into Brownian motion by introducing the notion of a first passage time, which is a particular example of a stopping time. On the other hand, the inclusion of a totally or partially reflecting boundary requires a modification of the stochastic process itself. Mathematically speaking, one can construct so-called reflected Brownian motion in terms of a boundary local time, which characterizes the amount of time that a Brownian particle spends in the neighborhood of a point on a totally reflecting boundary \cite{Levy39,McKean75}. The resulting stochastic differential equation, also known as the stochastic Skorokhod equation \cite{Freidlin85}, can then be extended to take into account chemical reactions, thus providing a probabilistic implementation of the Robin boundary condition \cite{Papanicolaou90,Milshtein95}. A simpler conceptual framework for understanding partially reflected Brownian motion is to model diffusion as a discrete-time random walk on a hypercubic lattice ${\mathbb Z}^d$ with lattice spacing $a$. At a bulk site, a particle jumps to one of the neighboring sites with probability $1/2d$, whereas at a boundary site, it either reacts with probability $q=(1+\xi/a)^{-1}$ or return to a neighboring bulk site with probability $1-q$. Since the random jumps are independent of the reaction events, it follows that the random number of jumps $\widehat{N}$ before a reaction occurs is given by a geometric distribution: $\P[\widehat{N}=n]=q(1-q)^n$, integer $n\geq 0$. In particular, $\E[\widehat{N}]=(1-q)/q=\xi/a$. Introducing the rescaled random variable $\widehat{\ell }=a\widehat{N}$, one finds that \cite{Grebenkov03,Grebenkov19a}
 \begin{align*}
  \P[\widehat{\ell }\geq \ell]&=\P[\widehat{N}\geq \ell/a]=(1-q)^{\ell/a} =(1+a/\xi)^{-\ell/a}\\&\underset{a\rightarrow 0}\rightarrow \e^{-\ell/\xi}.
  \end{align*}
 That is, for sufficiently small lattice spacing $a$, a reaction occurs (the random walk is terminated) when the random number of realized jumps from boundary sites, multiplied by $a$, exceeds an exponentially distributed random variable (stopping local time) $\widehat{\ell}$ with mean $\xi$. Assuming that a partially reflected random walk on a lattice converges to a well-defined continuous process in the limit $a\rightarrow 0$ (see Refs. \cite{Papanicolaou90,Milshtein95}), one can define partially reflected Brownian motion as reflected Brownian motion stopped at the random time \cite{Grebenkov06,Grebenkov07,Grebenkov19a}
 \begin{equation*}
 {\mathcal T}=\inf\{t>0:\ \ell_t >\widehat{\ell}\},
 \end{equation*}
where $\ell_t$ is the local time of the reflected Brownian motion. The latter is the continuous analog of the rescaled number of surface encounters ($a\widehat{N}$), and $\P[\widehat{\ell}>\ell]=\e^{-\ell/\xi}$. The reaction length $\xi$ thus parameterizes the stochastic process. (Note that it is also possible to construct more general partially reflecting diffusion processes by considering the continuous limit of more general Markovian jump processes \cite{Singer08}.)

 One major advantage of the above formulation of partially reflected Brownian motion is that it provides a theoretical framework for investigating more general diffusion-mediated surface phenomena \cite{Grebenkov19b,Grebenkov20,Grebenkov21}. In particular, by considering the joint probability density $P(\x,\ell,t)$ for the pair $(\X_t,\ell_t)$, where $\X_t$ is the particle position at time $t$ and $\ell_t$ is the boundary local time, one can analyze the bulk dynamics in a domain with perfectly reflecting boundaries and then incorporate the effects of surface reactions via an appropriate stopping condition for the boundary local time. In particular, the probability density $p(\x,t)$ for partially reflected Brownian motion can be expressed as the Laplace transform of a propagator $P$:
 \[p(\x,t)=\int_0^{\infty} \e^{-\gamma \ell}P(\x,\ell,t)d\ell,\]
 where $\gamma=\xi^{-1}=\kappa_0/D$. 
 This so-called encounter-based approach allows one to go beyond the case of constant reactivity (Robin boundary conditions) by considering more general probability distributions $\Psi(\ell) = \P[\ellh>\ell]$ for the stopping local time $\ellh$ and setting \cite{Grebenkov19b,Grebenkov20,Grebenkov21}
  \[p(\x,t)=\int_0^{\infty} \Psi(\ell)P(\x,\ell,t)d\ell.\]
For example, reaction rates could depend on the number of encounters between the particle and surface. 
 The separation of the bulk dynamics from surface reactions means that all of the geometrical aspects of the diffusion process are disentangled from the reaction kinetics. Geometrical features include the structure of both reactive and non-reactive surfaces. In the case of the narrow capture problem in a bounded domain $\Omega$, the exterior boundary of the domain, $\partial \Omega$, would correspond to a non-reactive surface, say, while the reactive surfaces would be given by the interior target boundaries. In the case of small targets, matched asymptotic methods provide a way of further separating geometrical effects. That is, the bulk dynamics is partitioned into an outer solution that depends on the exterior boundary and an inner solution that depends on the geometry of the targets. 
 
 In this paper, we analyze the 3D narrow capture problem for a $N$ small spherical targets with partially reactive boundary surfaces. (For simplicity, we consider the unbounded domain $\Omega=\R^3$, but see the discussion in Sect. V.) We proceed by combining the encounter-based approach to diffusion-mediated surface reactions \cite{Grebenkov19b,Grebenkov20,Grebenkov21} with matched asymptotic methods \cite{Bressloff21b}. We begin by considering the narrow capture problem for reactive surfaces with classical Robin boundary conditions, see Sect. II. Working in Laplace space, we construct an inner solution around each target, and then match it with an outer solution in the bulk. This yields an asymptotic expansion of the Laplace transformed flux into each reactive surface in powers of $\epsilon$, where $\epsilon$ is the non-dimensionalized target size. The Laplace transformed fluxes are then used to determine the splitting probabilities in the small-$s$ limit, where $s$ is the Laplace variable. In Sect. III we briefly summarize the encounter-based formulation of diffusion-mediated surface reactions developed in Ref. \cite{Grebenkov20}. In particular, we define the boundary local time $\ell_t$ for diffusion in a domain $\R^3\backslash \calU$ with a perfectly reflecting boundary $\partial \calU$ and write down the BVP for the associated propagator. It turns out that for the narrow capture problem, it is more convenient to work directly with the BVP rather then using the spectral decomposition of the propagator and the so-called Dirichlet-to-Neumann operator \cite{Grebenkov19a,Grebenkov19b,Grebenkov20}. In Sect. IV we use matched asymptotics to analyze the corresponding propagator BVP for the narrow capture problem, in which the reactive boundaries of the  targets are replaced by totally reflecting boundaries. This then allows us to incorporate generalized surface reactions by considering an appropriately defined distribution $\Psi(\ell)$ of stopping local times. We thus obtain an asymptotic expansion of the inner solution for the Laplace transformed probability density and the corresponding target fluxes. We also show that our results for Robin boundary conditions in Sect. II are recovered
 in the special case $\Psi(\ell)=\e^{-\gamma \ell }$. Finally, we illustrate the theory in Sect. V by exploring how the leading-order contributions to the splitting probabilities depend on the choice of surface reactions. In particular, we show that there is an effective renormalization of the target radius of the form $\rho\rightarrow \rho-\widetilde{\Psi}(1/\rho)$, where $\widetilde{\Psi}$ is the Laplace transform of the stopping local time distribution.

\setcounter{equation}{0} 
\section{Narrow capture problem: Robin boundary conditions}

\begin{figure*}[t!]
\centering
\includegraphics[width=11cm]{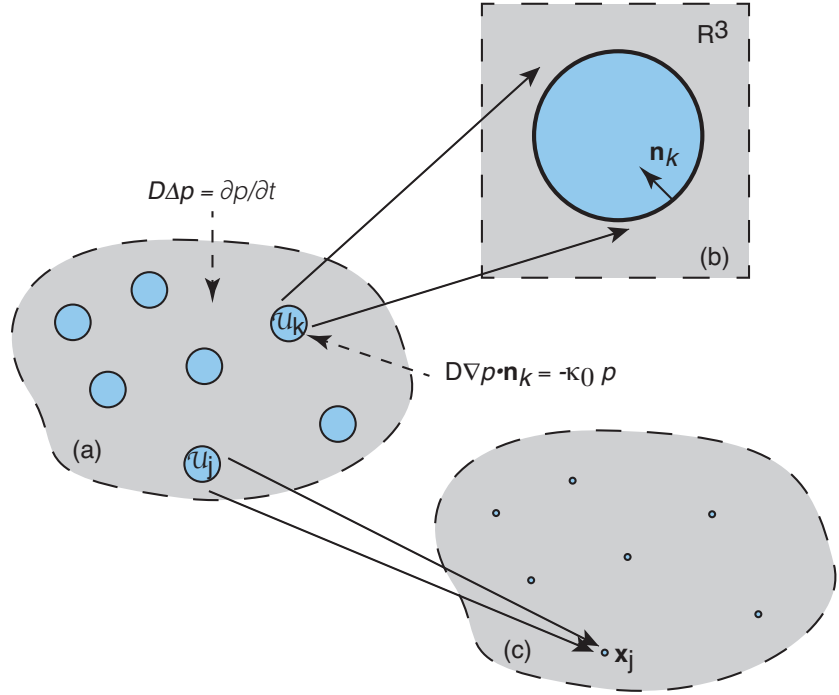} 
\caption{Brownian particle in a singularly perturbed domain. (a) A particle diffuses in the domain $\Omega=\R^3\backslash \cup_{j=1}^N\calU_j$ exterior to $N$ targets $\calU_j$, $j=1,\ldots,N$ whose boundaries $\partial \calU_i$ are partially absorbing. [Diagram is not to scale -- the radii of the targets are at least an order of magnitude smaller than the distances between the targets. (b) Construction of the inner solution in terms of stretched coordinates $\y=\epsilon^{-1}(\x-{\x}_i)$, where ${\x}_i$ is the center of the $i$-th target. The rescaled radius is $\rho_i$. (c) Construction of the outer solution. Each target is shrunk to a single point. The outer solution can be expressed in terms of the modified Helmholtz Green's function and then matched with the inner solution around each target.}
\label{fig1}
\end{figure*}

Consider a set of $N$ small partially absorbing targets $\calU_k\subset \R^3$, $k=1,\ldots,N$, see Fig. \ref{fig1}. Each target is assumed to have a volume $|\calU_j|\sim \epsilon^3 L^3$ with $\calU_j\rightarrow \x_j\in \R^3$ uniformly as $\epsilon \rightarrow 0$, $j=1,\ldots,N$. Here $L$ is the minimum separation between the targets.
For concreteness we will take each target to be a sphere of radius $r_j=\epsilon \rho_j$. Thus $\calU_i=\{\x \in \R^3, \ |\x-\x_i|\leq \epsilon \rho_i\}$.
Let $p(\x,t|\x_0)$ be the probability density that at time $t$ a particle is at $\X(t)=\x$, having started at position $\x_0$. Setting $\bigcup_{j=1}^N \calU_k=\calU_a\subset \R^3$, we have
\begin{subequations} 
\label{master}
\begin{align}
	\frac{\partial p(\x,t|\x_0)}{\partial t} &= D\nabla^2 p(\x,t|\x_0), \ \x\in \R^3\backslash \calU_a,\\
	& p(\x,t|\x_0)\rightarrow 0, \ |\x|\rightarrow \infty,\\ 
	D\nabla p(\x,t|\x_0) \cdot \n_k&=-\kappa_0 p(\x,t|\x_0),\  \x\in \partial\calU_k,
	\end{align}
	\end{subequations} 
together with the initial condition $p(\x,t|\x_0)=\delta(\x-\x_0)$. Here ${\bf n}_k$ is the unit normal into the surface $\partial \calU_k$. Eq. (\ref{master}b) is a Robin boundary condition with the constant reactivity parameter $\kappa_0$ having units $m/s$ \cite{Collins49}. Dirichlet and Neumann boundary conditions are recovered in the limits $\kappa_0\rightarrow \infty$ and $\kappa_0 \rightarrow 0$, respectively.

\subsection{Matched asymptotics}

In order to calculate various quantities of interest, it is more convenient to work in Laplace space:
\begin{subequations} 
\label{masterLT}
\begin{align}
	&D\nabla^2 \p(\x,s|\x_0) -s\p(\x,s|\x_0)= -\delta(\x-\x_0) , \ \x\in \R^3\backslash \calU_a,\\
	& \p_{\infty}(\x,s|\x_0)\rightarrow 0, \ |\x|\rightarrow \infty,\\ 
	 & D \nabla \p(\x,s|\x_0)\cdot \n_k= -\kappa_0 \p(\x,s|\x_0),\ \x\in \partial\calU_k.
	\end{align}
\end{subequations}
Let $\p_{\infty}(\x,s|\x_0)$ denote the solution in the case of totally absorbing targets, which corresponds to taking the limit $\kappa_0\rightarrow \infty$ in Eqs. (\ref{masterLT}):
\begin{subequations} 
\label{masterLTinf}
\begin{align}
	&D\nabla^2 \p_{\infty}(\x,s|\x_0) -s\p_{\infty}(\x,s|\x_0)= -\delta(\x-\x_0) \nonumber \\
	&\mbox{ for } \x\in \R^3\backslash \calU_a,\\
	& \p_{\infty}(\x,s|\x_0)\rightarrow 0, \ |\x|\rightarrow \infty,\\ &  \p_{\infty}(\x,s|\x_0)= 0,\ \x\in \partial\calU_k.
	\end{align}
\end{subequations}
Eqs. (\ref{masterLTinf}) define a BVP that has previously been solved using matched asymptotics and Green's function methods \cite{Cheviakov11,Coombs15,Bressloff21b}. Similar methods can be used to solve the full BVP (\ref{masterLT}), by matching appropriate
`inner' and `outer' asymptotic expansions in the
limit of small target size $\varepsilon\to 0$, see Figs. \ref{fig1}(b,c). However, given that $\p_{\infty}$ is known, it is more convenient to decompose the solution for finite $\kappa_0$ as
\begin{equation}
\p(\x,s|\x_0)=\p_{\infty}(\x,s|\x_0)+\u(\x,s|\x_0),
\end{equation}
with
\begin{subequations} 
\label{masterLTu}
\begin{align}
	&D\nabla^2 \u(\x,s|\x_0) -s\u(\x,s|\x_0)= 0 , \ \x\in \R^3\backslash \calU_a,\\
	 & D \nabla \u(\x,s|\x_0)\cdot \n_k+\kappa_0 \u(\x,s|\x_0)\nonumber \\
	&\quad =-D\nabla \p_{\infty}(\x,s|\x_0),\ \x\in \partial\calU_k.
	\end{align}
\end{subequations}
We then consider asymptotic expansions of $\u$.

In the outer region, $\u(\x,s|x_0)$ is expanded as
\[\u(\x,s|\x_0)\sim \epsilon \u_1(\x,s|\x_0)+\epsilon^2 \u_2(\x,s|\x_0)+ \ldots
\]
such that (for $m\geq 0$)
\begin{subequations}
\label{asym1}
\begin{align}
D\nabla^2 \u_m-s\u_m&=0,\quad \x\in \R^3\backslash \{\x_1,\ldots,\x_N\},
\end{align}
\end{subequations}
together with certain singularity conditions as $\x\rightarrow \x_j$, $j=1,\ldots,N$. The latter are determined by matching to the inner solution.

Next we introduce stretched coordinates ${\bf y}=\epsilon^{-1}(\x-\x_j)$ around the $j$th target and take
$\v(\y,s|\x_0)=\u(\x,s|\x_0)$ to be the corresponding inner solution. Eqs. (\ref{masterLTu}) imply 
\begin{subequations}
\label{noo}
\begin{align}
	&D\nabla_{\y}^2 \v(\y,s|\x_0) -s\epsilon^2 \v(\y,s|\x_0)= 0, \ |\y|>\rho_j,\\
	&D\nabla_{\y} \v(\y,s|\x_0)\cdot \n_j +\epsilon \kappa_0 \v(\y,s|\x_0)\nonumber \\
	&\quad =-D\nabla_{\y} \p_{\infty}(\y,s|\x_0)\cdot \n_j \,\ |\y|=\rho_j.
	\end{align}
	\end{subequations}
	The details of the analysis of the inner solution will now depend on how the reaction length $\xi=D/\kappa_0$ compares to the typical target size $\epsilon \bar{\rho}$, where $\bar{\rho}=N^{-1}\sum_{i=1}^N\rho_j$ for example \cite{Ward00}. We will focus on the regime $\xi \sim \epsilon \bar{\rho}$ by rescaling the reactivity according to $\kappa_0\rightarrow \kappa_0/\epsilon$. (Under this choice of scaling, we can recover the totally absorbing case by taking $\kappa_0 \rightarrow \infty$, that is, $\xi \rightarrow 0$. However, the totally reflecting case $\kappa_0\rightarrow 0$ is inaccessible.) Introduce a perturbation expansion of the inner solutions around the $j$-th target of the form
	\begin{align*}
	\v &\sim \v_0 + \epsilon \v_1 +O(\epsilon^2).
	\end{align*}
It remains to specify the corresponding asymptotic expansion of the totally absorbing solution. Let $G(\x,s|\x_0)$ denote the Green's function of the modified Helmholtz equation in $\R^3$:
\begin{equation}
G(\x,s|\x_0)=\frac{\e^{-\sqrt{s/D} |\x-\x_0|}}{4\pi D|\x-\x_0|}=\frac{1}{4\pi D|\x-\x_0|}+R(\x,s|\x_0),
\end{equation}
where $R$ is the regular part of $G$. 
Then
	\begin{equation}
	\label{pinf}
	\p_{\infty} \sim \p_{\infty,0} + \epsilon \p_{\infty,1} +O(\epsilon^2),
	\end{equation}
	with \cite{Bressloff21a}
	\begin{subequations}
	\label{ciao}
	\begin{align}
	\p_{\infty,0} &=G(\x_j,s|\x_0)\left (1-\frac{\rho_j}{|\y|}\right ),\\
	\p_{\infty,1}&=\bar{\chi}_j (s)\left (1-\frac{\rho_j}{|\y|}\right )\nonumber \\
	&\quad +\mbox{ first-order spherical harmonics}.
		\end{align}
(The explicit form of the first-order spherical harmonics is not needed here, since it does not contribute to the target flux.) The coefficient $\overline{\chi}_j$ is
\begin{equation}
\overline{\chi}_j(s) = -4\pi D\sum_{k=1}^N \rho_k  G_{k0}(s){\mathcal G}_{jk}(s) ,
\end{equation}
\end{subequations}
where $G_{k0}(s)=G(\x_k,s|\x_0)$ and
\begin{equation}
\label{calG}
{\mathcal G}_{ij}(s) =G(\x_i,s|\x_j) \mbox{ for } i\neq j, \quad {\mathcal G}_{ii}(s) = R(\x_i,s|\x_i).
\end{equation}

Substituting into Eqs. (\ref{noo}) leads to the following equations (assuming $s\ll 1/\epsilon$)
\begin{subequations} 
\label{stretch}
\begin{align}
&D\nabla_{\y}^2 \v_m(\y,s|\x_0) =0,\ |\y|>\rho_j,\quad m=0,1,\\ 
	&D\nabla_{\y}^2 \v_m(\y,s|\x_0) =s\epsilon^2 \v_{m-2}(\y,s|\x_0)= 0, \quad  m\geq 2,\\
	&D\nabla_{\y} \v_m(\y,s|\x_0)\cdot \n_j +\kappa_0 \v_m(\y,s|\x_0)\nonumber \\
	 &=-D\nabla_{\y} \p_{\infty,m}(\y,s|\x_0)\cdot \n_j, \,\ |\y|=\rho_j,\, m\geq 0.
	\end{align}
\end{subequations}
These are supplemented by far-field conditions obtained by matching with the near-field behavior of the outer solution. In order to perform the matching, it is necessary to consider the Taylor expansion of $\p_{\infty}$ near
the $j$-th target: 
\begin{align}
\p_{\infty}& \sim p_{\infty}(\x_j,s|\x_0) + \nabla_{\x} \p_{\infty}(\x,s|\x_0)\vert_{\x=\x_j} \cdot (\x-\x_j)\nonumber 
\\ &\sim \epsilon \nabla_{\x} \p_{\infty}(\x_j,s|\x_0) \cdot \y +\ldots,
\label{p0}
\end{align}
 since $p_{\infty}(\x_j,s|\x_0)=0$.

Let us begin with the leading order contribution to the inner solution. Matching the far-field behavior of $\v_0$ with the near-field behavior of $\p_{\infty}$ shows that
\begin{subequations}
\begin{align}
& \nabla_{\bf y}^2 \v_0(\y,s|\x_0) =0,\  |\y|>1,\,  \v_0 \sim 0 \mbox{ as } |\y|\to \infty;\\ 
&  D\nabla_{\y} \v_0(\y,s|\x_0)\cdot \n_j  +\kappa_0 \v_0(\y,s|\x_0)\nonumber \\
&\quad =-D\nabla_{\y} \p_{\infty,0}(\y,s|\x_0)\cdot \n_j,\quad   \ |\y|=\rho_j. 
\end{align}
\end{subequations}
In the case of a spherical target of radius $\rho_j$, we have
\begin{equation}
\label{inn0}
\v_0 = \frac{G_{j0}(s))}{1+\gamma \rho_j} \frac{\rho_j}{|\y|},\quad \gamma =\frac{\kappa_0}{D}.
\end{equation}
It follows that $\p_1$ satisfies Eq. (\ref{asym1}) together with the singularity condition
\[\u_1(\x,s|\x_0)\sim  \frac{1}{1+\gamma \rho_j}\frac{G_{j0}(s)\rho_j}{|\x-\x_j|} \quad \mbox{as } \x\rightarrow \x_j.\]
In other words, $\u_1$ satisfies the inhomogeneous equation
\begin{subequations}
\begin{align}
 D\nabla^2 \u_1-s\u_1&=-{4\pi D} \sum_{j=1}^N\frac{G_{j0}(s)\rho_j}{1+\gamma \rho_j }\delta(\x-\x_j),\, \x\in \R^3.
 \label{asym2}
\end{align}
\end{subequations}
This can be solved in terms of the modified Helmholtz Green's function:
\begin{equation}
\label{q1}
\u_1(\x,s|\x_0)= {4\pi}  D\sum_{j=1}^N\frac{G_{j0}(s)\rho_j}{1+\gamma \rho_j}G(\x,s|\x_j).
\end{equation}
 We now match the far-field behavior of $\v_1$ with the $O(\epsilon)$ term in the expansion of $\p_{\infty}$, see Eq. (\ref{p0}), together with the non-singular near-field behavior of $\u_1 $ around the $j$-th target:
\begin{align}
\v_1(\y,s|\x_0)& \rightarrow  \nabla_{\x} G(\x_j,s|\x_0) \cdot \y \nonumber \\
&\quad +4\pi D\sum_{k=1}^N\frac{G_{k0}(s)\rho_k}{1+\gamma \rho_k}{\mathcal G}_{jk}(s) 
\end{align}
as $ |\y|\rightarrow \infty$. We thus obtain a solution of the form
\begin{align}
\v_1(\y,s|\x_0) &=\overline{\chi}_j' \left (1-\frac{\rho_j}{|\y|}\right )+(\overline{\chi}_j' +\chi_j)\frac{\rho_j}{(1+\gamma \rho_j)|\y|}\nonumber \\
&\quad +\mbox{ first-order spherical harmonics},
\end{align}
with
\begin{equation}
\overline{\chi}_j' (s)\equiv 4\pi D\sum_{k=1}^NG_{k0}(s) \frac{\rho_k}{1+\gamma \rho_k}  {\mathcal G}_{jk}(s) .
\end{equation}

Combining our various results, the full inner solution is
\begin{align}
\p(\y,s|\x_0)&=\p_{\infty,0}(\y,s|\x_0)+\v_0(\y,s|\x_0)\nonumber \\
&\quad +\epsilon [\p_{\infty,1}(\y,s|\x_0)+\v_1(\y,s|\x_0)]+O(\epsilon^2)\nonumber \\
&=G_{j0}(s)\left (1-\frac{\rho_j}{|\y|}-\frac{\rho_j}{(1+\gamma \rho_j)|\y|}\right )\\
&\quad +\epsilon (\overline{\chi}_j' +\chi_j)\left (1-\frac{\rho_j}{|\y|}+\frac{\rho_j}{(1+\gamma \rho_j)|\y|}\right ).\nonumber 
\end{align}
Introducing the renormalized target radius
\begin{equation}
\rho_j^{\gamma} =\rho_j-\frac{\rho_j}{1+\gamma \rho_j},
\end{equation}
we can write the inner solution as
\begin{align}
\p(\y,s|\x_0)&=G_{j0}(s)\left (1-\frac{\rho_j^{\gamma}}{|\y|}\right ) +\epsilon \overline{\chi}_j^{\gamma}(s) \left (1-\frac{\rho_j^{\gamma}}{|\y|}\right )\nonumber \\
&\quad +O(\epsilon^2),
\label{inner}
\end{align}
where
\begin{equation}
\overline{\chi}_j^{\gamma}(s) = -4\pi D\sum_{k=1}^NG_{k0}(s) \rho_k^{\gamma}  {\mathcal G}_{jk}(s) .
\end{equation}

\subsection{The flux into a target}

The probability flux into the $j$-th target at time $t$ is 
\begin{align}
\label{J}
	J_j(\x_0,t)&=-D \int_{\partial \calU_j} \nabla p(\x,t|\x_0)\cdot \n_j d\sigma
	\end{align}
for $j = 1,\ldots,N$, where $d\sigma$ is the surface measure. Having obtained an $\epsilon$ expansion of the inner solution in stretched coordinates, we can determine a corresponding expansion of the Laplace-transformed flux through the $j$th target by substituting Eq. (\ref{inner}) into the Laplace transform of Eq. (\ref{J}):
\begin{align}
\widetilde{J}_j(\x_0,s)&= -D\epsilon^2 \int_{|\y|=\rho_j} \nabla \p(\y,s|\x_0) \cdot \n_jd\sigma_{\y} \\
&=D\rho_j^2 \int_0^{2\pi}\int_0^{\pi} \left .\frac{\partial}{\partial r}\right |_{r=\rho_j}\p(\y,s|\x_0)\sin \theta d\phi d\theta .\nonumber 
\end{align}
We thus obtain the result
\begin{align}
\label{JLT1}
\widetilde{J}_j(\x_0,s)
&\sim 4\pi \epsilon D\rho_j^{\gamma}\bigg (G_{j0}(s)\\
&\quad -4\pi \epsilon D\sum_{k=1}^NG_{k0}(s)\rho_k^{\gamma}{\mathcal G}_{jk}(s)\bigg )
+O(\epsilon^3).\nonumber \end{align}
  
One application of diffusion to a target in an unbounded domain is calculating the effective Smoluchowski reaction rate in terms of the steady-state flux into the target. 
Suppose that there is a continuous concentration $c(\x,t)$ of non-interacting diffusing particles with background concentration $c_0$, that is, $c(\x,t)\rightarrow c_0$ as $|\x|\rightarrow \infty$. The steady-state flux into the $j$th target is obtained by integrating over the initial position $\x_0$ according to
\begin{equation}
J_j=c_0 \lim_{s\rightarrow 0}s\int_{\R^3}\widetilde{J}_j(\x_0,s) d\x_0,
\end{equation}
with $\widetilde{J}_j(\x_0,s) $ given by Eq. (\ref{JLT1}). Using the fact that
\[\int_{\R^3}G(\x,s|\x_0)d\x_0=\frac{1}{s},\]
it follows that to leading order
\begin{equation}
J_j\approx 4\pi c_0 \epsilon D\rho_j^{\gamma}= \frac{4\pi c_0  D r_j}{1+\epsilon D/\kappa_0 r_j},
\end{equation}
where $r_j=\epsilon \rho_j$ is the target radius. This recovers the modified Smoluchowksi reaction rate obtained by Collins and Kimball for a partially reactive spherical surface with reactivity $\kappa_0/\epsilon$ \cite{Collins49}. In particular, note that one way to interpret the effect of imperfect reactivity is that the effective traget size is reduced according to
\[r_j\rightarrow  \frac{ r_j}{1+\epsilon D/\kappa_0 r_j},\]
thus making it more difficult for a diffusing molecule to encounter it. This result generalizes to other types of diffusion-mediated surface reactions, see Sect. V.

Another quantity of interest is the splitting probability that the particle is eventually captured by the $k$-th target :
\begin{equation}
\label{split}
\pi_k(\x_0)=\int_0^{\infty}J_k(\x_0,t')dt' =\widetilde{J}_k(\x_0,0).
\end{equation}
Introduce the survival probability that the particle hasn't been absorbed by a target in the time interval $[0,t]$, having started at $\x_0$:
\begin{equation}
\label{S1}
S(\x_0,t)=\int_{\R^3\backslash \calU_a}p(\x,t|\x_0)d\x.
\end{equation}
Differentiating both sides of this equation with respect to $t$ and using Eqs. (\ref{master}) implies that
\begin{align}
\label{Q2}
&\frac{\partial S(\x_0,t)}{\partial t}=D\int_{\R^3\backslash \calU_a}\nabla\cdot \nabla p(\x,t|\x_0)d\x \\
&=D\sum_{k=1}^N \int_{ \partial \calU_k}\nabla p(\x,t|\x_0)\cdot \n d\sigma =-\sum_{k=1}^NJ_k(\x_0,t).\nonumber
\end{align}
Laplace transforming Eq. (\ref{Q2}) and noting that $S(\x_0,0)=1$ gives
\begin{equation}
\label{QL}
s\widetilde{S}(\x_0,s)-1=- \sum_{k= 1}^N \widetilde{J}_k(\x_0,s).
\end{equation}
An asymptotic expansion of the splitting probability $\pi_j(\x_0)$ defined in Eq. (\ref{split}) can now be obtained by taking the limit $s\rightarrow 0$ in Eq. (\ref{JLT1}):
\begin{align}
 &\pi_j(\x_0)=\lim_{s\rightarrow 0}\widetilde{J}_j(\x_0,s) \\
 &= \epsilon  \rho_j^{\gamma}\left [\frac{1}{|\x_j-\x_0|}- \epsilon \sum_{k\neq j} \frac{\rho_k^{\gamma}}{|\x_k-\x_0||\x_k-\x_j|}\right ]    +O(\epsilon^3),\nonumber
\label{split2}
\end{align}
since $R(\x_j,0|\x_j)=0$.

\setcounter{equation}{0}
\section{Boundary local time and the propagator} In this section we introduce the encounter-based formulation of diffusion-mediated surface reactions developed in Ref. \cite{Grebenkov20}. We begin by giving a brief heuristic definition of the boundary local time. For more rigorous treatments see Refs. \cite{Levy39,McKean75,Freidlin85}. Consider the Brownian motion $X_t\in \R$, and let ${\mathcal  T}(A,t)$ denote the occupation time of the set $A\subset \R$ during the time interval $[0,t]$:
\begin{equation}
{\mathcal  T}(A,t)=\int_{0}^tI_A(X_{\tau})d\tau.
\end{equation}
Here $I_A(x)$ denotes the indicator function of the set $A\subset \R$, that is, $I_A(x)=1$ if $x\in A$ and is zero otherwise.
From the definition of the occupation time, the local time density ${\mathcal  T}(a,t)$ at a point $a\in \R$ is defined by setting $A=[a-h,a+h]$ and taking
\begin{equation}
{\mathcal  T}(a,t)=\underset{\epsilon \rightarrow 0^+}\lim \frac{1}{2h}\int_{0}^tI_{[a-h,a+h]}(X_s)ds.
\end{equation}
We thus have the following formal representation of the local time density:
\begin{equation}
\label{ugate}
{{\mathcal  T}(a,t)=\int_{0}^t \delta(X_{\tau}-a)d\tau,}
\end{equation}
where ${\mathcal  T}(a,t)da$ is the amount of time the Brownian particle spends in the infinitesimal interval $[a,a+da]$. Note, in particular, that
\[\int_{-\infty}^{\infty} {\mathcal  T}(a,t)da=\int_{-\infty}^{\infty}\, \int_0^t\delta(X_{\tau}-a)d\tau da =\int_0^td\tau = t.\]
As we mentioned in the introduction, local time plays an important role in the pathwise formulation of reflected Brownian motion \cite{McKean75}. For the sake of illustration, consider a Wiener process confined to the interval $[0,L]$ with reflecting boundaries at $x=0,L$. Sample paths are generated from the stochastic differential equation
\begin{equation}
dX(t)=\sqrt{2D} dW(t)+Dd{\mathcal  T}(0,t)-D d{\mathcal  T}(L,t),
\end{equation}
where ${\mathcal  T}(x,t)$ is given by Eq. (\ref{ugate}) so that, formally speaking,
\[d{\mathcal  T}(0,t)=\delta(X_t)dt,\quad d{\mathcal  T}(L,t)=\delta(X_t-L)dt.\]
In other words, each time the Brownian particle hits the end at $x=0$ ($x=L$) it is given an impulsive kick to the right (left). 

Following Ref. \cite{Grebenkov20}, we now define the boundary local time for diffusion in $\R^3\backslash \calU$ for a single obstacle with a totally reflecting surface $\partial \calU$:
\begin{equation}
\ell_t=\lim_{h\rightarrow 0} \frac{D}{h} \int_0^t\Theta(h-\mbox{dist}(\X_{\tau},\partial \calU))d\tau,
\end{equation}
where $\Theta$ is the Heaviside function. Note that $\ell_t$ has units of length due to the additional factor of $D$.
Given the definition of the boundary local time $\ell_t$ for reflected Brownian motion at a surface $\partial \calU$, one can construct partially reflected Brownian motion by introducing the stopping time \cite{Grebenkov06,Grebenkov07,Grebenkov20}
\begin{equation}
{\mathcal T}_{\gamma}=\inf\{t>0:\ \ell_t >\widehat{\ell}\},
 \end{equation}
 with $\widehat{\ell}$ an exponentially distributed random variable that represents a stopping local time. That is, $\P[\widehat{\ell}>\ell]=\e^{-\gamma\ell}$
with $\gamma=\xi^{-1}=\kappa_0/D$. Let $p(\x,t|\x_0) $ be the probability density for a Brownian particle to be at position $\x\in \R^3\backslash\calU$ at time $t$, having started at $\x_0$ and given a constant inverse reaction length $\gamma$. Then
\begin{subequations} 
\label{master2}
\begin{align}
	\frac{\partial p(\x,t|\x_0)}{\partial t} &= D\nabla^2 p(\x,t|\x_0), \ \x\in \R^3\backslash \calU,\\
\nabla p(\x,t|\x_0) \cdot \n&=-\gamma p(\x,t|\x_0),\  \x\in \partial \calU,\\p(\x,0|\x_0)&=\delta(\x-\x_0).
	\end{align}
	\end{subequations} 
More precisely, $p$ is 
the probability density of a particle that hasn't yet undergone a surface reaction:
\[p(\x,t|\x_0)d\x=\P[\X_t \in (\x,\x+d\x), \ t < {\mathcal T}_{\gamma}|\X_0=\x_0].\]
Given that $\ell_t$ is a nondecreasing process, the condition $t < {\mathcal T}_{\gamma}$ is equivalent to the condition $\ell_t <\widehat{\ell}$. This implies that \cite{Grebenkov20}
\begin{align*}
&p(\x,t|\x_0)d\x=\P[\X_t \in (\x,\x+d\x), \ \ell_t < \widehat{\ell} |\X_0=\x_0]\\
&=\int_0^{\infty} d\ell \ \gamma\e^{-\gamma\ell}\P[\X_t \in (\x,\x+d\x), \ \ell_t < \ell |\X_0=\x_0]\\
&=\int_0^{\infty} d\ell \ \gamma \e^{-\gamma\ell}\int_0^{\ell} d\ell' [P(\x,\ell',t|\x_0)d\x],
\end{align*}
where $P(\x,\ell,t|\x_0)$ is the joint probability of the position $\X_t$ and boundary local time $\ell_t$ of reflected Brownian motion. We shall refer to $P$ as the propagator. (Note that Grebenkov refers to the density $p$ as the conventional propagator and denotes it by the symbol $G$ \cite{Grebenkov20,Grebenkov21}. The corresponding joint probability density $P$ is called the full propagator. In our paper we use $G$ to denote a Neumann Green's function and simply refer to $P$ as the propagator of reflected Brownian motion.) Using the identity
\[\int_0^{\infty}d\ell \ f(\ell)\int_0^{\ell} d\ell' \ g(\ell')=\int_0^{\infty}d\ell' \ g(\ell')\int_{\ell'}^{\infty} d\ell \ f(\ell)\]
for arbitrary integrable functions $f,g$, it follows that
\begin{equation}
p(\x,t|\x_0,\gamma)=\int_0^{\infty} \e^{-\gamma\ell}P(\x,\ell,t|\x_0)d\ell.
\end{equation}
Since the Robin boundary condition maps to an exponential law for the stopping local time $\widehat{\ell}_t$, the probability density $p(\x,t|\x_0,\gamma)$ can be expressed in terms of the Laplace transform of the propagator $P(\x,\ell,t|\x_0)$ with respect to the local time $\ell$.

The crucial observation is that one is free to change the probability distribution of the stopping local time $\widehat{\ell}$. Given some distribution $\Psi(\ell)= \P[\widehat{\ell}>\ell]$, one can define a generalized partially reflecting Brownian motion whose probability density is given by \cite{Grebenkov20}
\begin{equation}
\label{ppsi}
p(\x,t|\x_0)=\int_0^{\infty}\Psi(\ell) P(\x,\ell,t|\x_0)d\ell.
\end{equation}
In other words, the encounter-based formulation provides a framework for exploring a range of surface reaction mechanisms that go well beyond the constant reactivity case and exponential law $\Psi(\ell)=\e^{-\gamma \ell}$ associated with the Robin boundary condition. For example, one could consider a reactivity $\kappa(\ell)$ that depends on the local time $\ell$ (or the rescaled number of surface encounters). The corresponding distribution of the stopping local time $\widehat{\ell}$ would then be
\begin{equation}
\label{kaell}
\Psi(\ell)=\exp\left (-\frac{1}{D}\int_0^{\ell}\kappa(\ell')d\ell'\right ).
\end{equation}
However, for a more general surface reaction mechanism, one cannot calculate the probability density $p(\x,t|\x_0) $ by solving a BVP, since the Robin boundary condition no longer holds. This motivates the construction of the propagator $P(\x,\ell,t|\x_0)$, which is carried out in Ref. \cite{Grebenkov20} using a non-standard integral representation of the probability density $p(\x,t|\x_0)$ and spectral properties of the so-called Dirichlet-to-Neumann operator. In this paper it will be more convenient to work directly with the BVP for the propagator. In the case of a partially reactive boundary $\partial \calU$, the BVP takes the following form \cite{Grebenkov20}:
\begin{subequations} 
\label{prop0}
\begin{align}
&	\frac{\partial P(\x,\ell,t|\x_0)}{\partial t} = D\nabla^2 P(\x,\ell,t|\x_0), \ \x\in \R^3\backslash \calU\\
&-D\nabla P(\x,\ell,t|\x_0) \cdot \n=-D\nabla p_{\infty}(\x,t|\x_0)\cdot \n \ \delta(\ell) \nonumber \\
& \quad +D\frac{\partial}{\partial \ell} P(\x,\ell,t|\x_0),\  \x\in \partial \calU,\\
& P(\x,\ell=0,t|\x_0)=-\nabla p_{\infty}(\x,t|\x_0)\cdot \n ,\ \x\in \partial \calU,\\
&\lim_{\ell \rightarrow \infty}P(\x,\ell,t|\x_0)=0,\\ &P(\x,\ell,0|\x_0)=\delta(\x-\x_0)\delta(\ell),\quad \x \in \R^3\backslash \calU,
	\end{align}
	\end{subequations} 
	where $p_{\infty}$ is the probability density for a totally absorbing surface.
Note that multiplying the boundary condition (\ref{prop0}b) by $\e^{-\gamma\ell}$, integrating with respect to $\ell\in [0,\infty)$, and using integration by parts combined with Eq. (\ref{prop0}c) recovers the standard Robin boundary condition for $p(\x,t|\x_0)$. In appendix A we present an alternative derivation of Eq. (\ref{prop0}) that is based on a Feynman-Kac equation, see Ref. \cite{Bressloff22}.

\setcounter{equation}{0}

\section{Narrow capture problem: generalized surface reactions}

In this section we use the encounter-based formulation \cite{Grebenkov20} to analyze the narrow capture problem shown in Fig. \ref{fig1} in the case of more general diffusion-mediated surface reactions. For simplicity, we take each target to have the same rule for surface reactions so that we only need to keep track of a single boundary local time that does not distinguish between targets
The BVP for the propagator of the system shown in Fig. \ref{fig1} can then be written down by analogy with Eq. (\ref{prop0}). Again it will be more convenient to work in Laplace space so that 
\begin{subequations} 
\label{prop}
\begin{align}
&	D\nabla^2 \PP(\x,\ell,s|\x_0)-s\PP(\x,\ell,s|\x_0)\nonumber \\
&\quad =-\delta(\x-\x_0)\delta(\ell), \ \x\in \R^3\backslash \calU_a,  \\
&-D\nabla \PP(\x,\ell,s|\x_0) \cdot \n_k=-D\nabla \p_{\infty}(\x,s|\x_0)\cdot \n_k \ \delta(\ell) \nonumber \\
&\quad +D\frac{\partial}{\partial \ell} \PP(\x,\ell,s|\x_0),\  \x\in \partial \calU_k,\\
&\left . \PP(\x,\ell,s|\x_0)\right |_{\ell=0}=-\nabla {\p}_{\infty}(\x,s|\x_0)\cdot \n_k ,\ \x\in \partial \calU_k,\\
&\lim_{\ell \rightarrow \infty}\PP(\x,\ell,s|\x_0)=0 .
	\end{align}
	\end{subequations} 
It is convenient to eliminate the terms involving Dirac delta functions by setting
\begin{equation}
\PP(\x,\ell,s|\x_0)=\p_{\infty}(\x,s|\x_0)\delta(\ell) +\U(\x,\ell,s|\x_0),
\end{equation}
with
\begin{subequations} 
\label{propU}
\begin{align}
&	D\nabla^2 \U(\x,\ell,s|\x_0)-s\U(\x,\ell,s|\x_0)=0, \ \x\in \R^3\backslash \calU_a,   \\
&-D\nabla \U(\x,\ell,s|\x_0) \cdot \n_k=D\frac{\partial}{\partial \ell} \U(\x,\ell,s|\x_0),\  \x\in \partial \calU_k,\\
&\left . \U(\x,\ell,s|\x_0)\right |_{\ell=0}=-\nabla {\p}_{\infty}(\x,s|\x_0)\cdot \n_k ,\ \x\in \partial \calU_k,\\
&\lim_{\ell \rightarrow \infty}\U(\x,\ell,s|\x_0)=0.
	\end{align}
	\end{subequations}

\subsection{Asymptotic expansion of the propagator}
 
Following along analogous lines to the asymptotic analysis of Sect. II, we separately consider outer and inner solutions for the propagator. In the outer region, $\PP(\x,\ell,s|x_0)$ is expanded as
\begin{align*}\PP(\x,\ell,s|\x_0)&\sim \p_{\infty}(\x,s|\x_0)\delta(\ell)+\U_0(\x,\ell,s|\x_0)\\
&\quad +\epsilon\U_1(\x,\ell,s|\x_0)+\epsilon^2 \U_2(\x,\ell,s|\x_0)+\ldots,
\end{align*}
where 
\begin{subequations} 
\label{propout}
\begin{align}
&	D\nabla^2 \U_m(\x,\ell,s|\x_0)-s\U_m(\x,\ell,s|\x_0)=0, \nonumber\\
&\quad   \x\in \R^3\backslash \{\x_1,\ldots,\x_N\},\\
&\lim_{\ell \rightarrow \infty}\U_m(\x,\ell,,s|\x_0)=0.
	\end{align}
	\end{subequations} 
Eqs. (\ref{propout}) are supplemented
by singularity conditions as $\x\rightarrow \x_j$, $j=1,\ldots,N$, which are determined by matching to the inner solution.

Next consider the inner solution around the $j$th target. Introduce the stretched coordinates ${\bf y}=\epsilon^{-1}(\x-\x_j)$ and $\hat{\ell}=\ell/\epsilon$, and take
$\V(\y,\ellh,s|\x_0)=\epsilon \U(\x,\ell,s|\x_0)$ to be the corresponding inner solution. Eqs. (\ref{propU}) then imply that
\begin{subequations} 
\begin{align}
	&D\nabla_{\y}^2 \V(\y,\ellh,s|\x_0) -s\epsilon^2 \V(\y,\ellh,s|\x_0)= 0, \ |\y|>\rho_j,\\
	&D\nabla_{\y} \V(\y,\ellh,s|\x_0) \cdot \n_j= -D\frac{\partial}{\partial \ellh} \V(\y,\ellh,s|\x_0),\  |\y|=\rho_j,\\
	& \V(\y,\ellh=0,s|\x_0)=-\nabla_{\y} {\p_{\infty}}(\y,s|\x_0)\cdot \n_j ,\ |\y|=\rho_j.
	\end{align}
	\end{subequations} 
The choice of scaling for $\ell$ is consistent with a reactivity of $O(1/\epsilon)$, as assumed in Sect. II.
Introducing a perturbation expansion of the inner solution around the $j$-th target of the form
	\begin{equation}
	\label{Vexp}
	\V \sim \V_0 + \epsilon \V_1 + \epsilon^2\V_2+O(\epsilon^3)
	\end{equation}
	then yields the following pair of equations for $m=0,1$:
\begin{subequations} 
\label{propin}
\begin{align}
&D\nabla_{\y}^2 \V_m(\y,\ellh,s|\x_0) =0,\ |\y|>\rho_j, \\ 
		&D\nabla_{\y} \V_m(\y,\ellh,s|\x_0) \cdot \n_j=-D\frac{\partial}{\partial \ellh} \V_m(\y,\ellh,s|\x_0),\  |\y|=\rho_j,\\
		& \V_m(\y,\ellh=0,s|\x_0)=-\nabla_{\y} \p_{\infty,m}(\y,s|\x_0)\cdot \n_j ,\ |\y|=\rho_j.
	\end{align}
\end{subequations}

Let us begin with the leading order contribution to the inner solution. Matching the far-field behavior of $\V_0$ with the near-field behavior of $\epsilon p_{\infty}\delta(\ell)$ (which is zero) shows that the solution to Eq. (\ref{propin}a) for $m=0$ is of the form
\begin{equation}
\label{ginn0}
\V_0(\y,\ellh,s|\x_0) = \frac{c_j(\ellh)}{|\y|}.
\end{equation}
Substituting into the boundary conditions (\ref{propin}b,c) implies that
\begin{equation}
\frac{dc_j(\ellh)}{d\ellh}+\rho_j^{-1}c_j(\ellh)=0.
\end{equation}
Hence, $c_j(\ellh)=c_j(0)\e^{-\ellh/\rho_j}$ and
\begin{align}
\V_0(\y,\ellh,s|\x_0)=\frac{c_j(0)\e^{-\ellh/\rho_j}}{|\y|} ,
\label{Q0}
\end{align}
with
\begin{align}
c_j(0)&=-\rho_j\nabla_{\y} {\p_{\infty,0}}(\y,s|\x_0)\cdot \n_j|_{|\y|=\rho_j}\\
&= \rho_jG_{j0}(s)\left . \frac{d}{d\rho}\left (1-\frac{\rho_j}{|\y|}\right )\right |_{|\y|=\rho_j}=G_{j0}(s).\nonumber 
\end{align}

Rewriting Eq. (\ref{Q0}) in terms of the original unstretched coordinates then determines the singularity condition for $\U_0$:
\[\U_0(\x,\ell,s|\x_0)\sim \frac{G_{j0}(s)}{|\x-\x_j|}   \e^{-\ell/r_j}  \quad \mbox{as } \x\rightarrow \x_j.\]
The solution of Eq. (\ref{propout}) for $m=0$ is thus given by
\begin{align}
\label{qq1}
&\U_0(\x,\ell,s|\x_0) ={4\pi}  D\sum_{j=1}^NG_{j0}(s)  \e^{-\ell/r_j} G(\x,s|\x_j),
\end{align}
where $r_j=\epsilon \rho_j$.
 We now match the far-field behavior of $\V_1$ with the $O(\epsilon)$ term in the expansion of $\p_{\infty}(\x,s|\x_0)$ about $\x_j$ (multiplied by $\delta(\ell)$) together with the non-singular near-field behavior of $\U_0 $ around the $j$-th target. This yields 
\begin{align}
\V_1(\y,\ellh,s|\x_0)&\rightarrow  \nabla_{\x} p_{\infty}(\x_j,s|\x_0) \cdot \y\ \delta(\ellh) \nonumber \\
&\quad +4\pi D\sum_{k=1}^NG_{k0}(s) \e^{-\ellh/ \rho_k} {\mathcal G}_{jk}(s) \end{align}
as $ |\y|\rightarrow \infty$, with $\calG_{ij}$ defined in Eq. (\ref{calG}).
Following the analysis of Sect. II, we obtain
the general solution 
\begin{align}
&\V_1(\y,\ellh,s|\x_0) 
= 4\pi D\left [\sum_{k=1}^NG_{k0}(s)  \e^{-\ellh/ \rho_k}  {\mathcal G}_{jk}(s) \right ]\left ( 1 - \frac{\rho_j}{|\y|} \right ) \nonumber \\
&\quad +\frac{a_j(\ellh)}{|\y|}+\mbox{ first-order spherical harmonics}.
\label{A1Q}
\end{align}
Substituting (\ref{A1Q}) into the boundary conditions (\ref{propin}b,c) implies that
\begin{equation}
\frac{da_j(\ellh)}{d\ellh}+\rho_j^{-1}a_j(\ellh)=4\pi D\sum_{k=1}^NG_{k0}(s)  \e^{-\ellh/ \rho_k}  {\mathcal G}_{jk}(s),
\end{equation}
with
\begin{align}
a_j(0)&=-\rho_j\nabla_{\y} {\p_{\infty,1}}(\y,s|\x_0)\cdot \n_j|_{|\y|=\rho_j}\nonumber \\
&= \rho_j\overline{\chi}_j(s) \left . \frac{d}{d\rho}\left (1-\frac{\rho_j}{|\y|}\right )\right |_{|\y|=\rho_j}=\overline{\chi}_j(s).
\end{align}
Hence,
\begin{align}
a_j(\ellh)&=\overline{\chi}_j(s)  \e^{-\ellh/\rho_j}+4\pi D G_{j0}(s) \ellh \, \e^{-\ellh/\rho_j}
{\mathcal G}_{jj}(s)\nonumber \\
&\quad +4\pi D\sum_{k\neq j}^NG_{k0}(s) \frac{\e^{-\ellh/\rho_k}-\e^{-\ellh/\rho_j}}
{\rho_j^{-1}-\rho_k^{-1}}{\mathcal G}_{jk}(s).
\end{align}
Combining our various results yields the $O(\epsilon)$ contribution to the inner solution for the propagator:
\begin{widetext}
\begin{align}
\V_1(\y,\ellh,s|\x_0) &=4\pi D\left [\sum_{k=1}^NG_{k0}(s)  \e^{-\ellh/ \rho_k}  {\mathcal G}_{jk}(s) \right ]\left ( 1 - \frac{\rho_j}{|\y|} \right ) +\frac{\overline{\chi}_j(s)}{|\y|}  \e^{-\ellh/\rho_j}+\frac{4\pi D}{|\y| }G_{j0}(s) \ellh \, \e^{-\ellh/\rho_j}
{\mathcal G}_{jj}(s)\nonumber \\
&\quad +\frac{4\pi D}{|\y|} \left [ \sum_{k=1}^NG_{k0}(s)  \left \{ \frac{\e^{-\ellh/\rho_k}-\e^{-\ellh/\rho_j}}
{\rho_j^{-1}-\rho_k^{-1}}  \right \} {\mathcal G}_{jk}(s) \right ]+\mbox{ first-order spherical harmonics}. 
\label{Q1}
\end{align}
\end{widetext}

Having obtained an asymptotic expansion of the inner solution of the propagator in Laplace space, we can use the transform (\ref{ppsi}) to construct the corresponding asymptotic expansion of the probability density. First, Laplace transforming Eq. (\ref{ppsi}) gives
\begin{equation}
\label{ppsi2}
\p(\x,s|\x_0)=\int_0^{\infty}\Psi(\ell) \PP(\x,\ell,s|\x_0)d\ell.
\end{equation}
The case of Robin boundary conditions is recovered by setting $\Psi(\ell)=\e^{-\gamma \ell}$ with $\gamma=\kappa_0/D$ and $\kappa_0$ a constant reactivity. Recall that in the analysis of Sect. II we rescaled $\kappa_0$ according to $\kappa_0 \rightarrow \kappa_0/\epsilon$ so that
$\Psi(\ell) =\e^{-\kappa_0 \ell/\epsilon D}= \e^{-q \ellh}$ with $\ellh=\ell/\epsilon$. Therefore, we take $\Psi=\Psi ( \ellh)$ and rewrite Eq. (\ref{ppsi2}) as
\begin{equation}
\label{ppsi3}
\p(\x,s|\x_0)=\epsilon \int_0^{\infty}\Psi(\ellh) \PP(\x, \epsilon \ellh,s|\x_0)d\ellh .
\end{equation}
Introducing stretched coordinates then gives the corresponding transform of the inner solution around each target:
\begin{align}
\label{ppsin}
&\p(\y,s|\x_0)\\
&\quad =\int_0^{\infty}\Psi(\ellh) [ \p_{\infty}(\y,s|\x_0)\delta(\ellh)+\V(\y,\ellh,s|\x_0])d\ellh .\nonumber
\end{align}
Substituting the asymptotic expansions (\ref{pinf}) and (\ref{Vexp}), and using the fact that the integral of an asymptotic expansion is also an asymptotic expansion, we have
	\begin{equation}
	\p \sim \p_0 + \epsilon \p_1 + \epsilon^2\p_2+O(\epsilon^3),
	\end{equation}
	with
\begin{align}
\label{ltp}
&\p_m (\y,s|\x_0)\\
&\quad =\int_0^{\infty}\Psi(\ellh) [ \p_{\infty,m}(\y,s|\x_0)\delta(\ellh)+\V_m(\y,\ellh,s|\x_0])d\ellh .\nonumber
\end{align}

Substituting Eqs. (\ref{Q0}) and (\ref{Q1}) into (\ref{ltp}) for $m=0$ and $m=1$, respectively, yields
\begin{align}
\p_0 (\y,s|\x_0) 
&=G_{j0}(s) \left [  1 -\frac{\rho_j}{|\y|}   +\frac{\widetilde{\Psi}(1/\rho_j)}{|\y|} \right ],
\label{QPsi0}
\end{align}
and
\begin{widetext}
\begin{align}
\p_1(\y,s|\x_0) &=\p_{\infty,1}(\y,s|\x_0)+4\pi D\left [\sum_{k=1}^NG_{k0}(s)  \widetilde{\Psi}(1/\rho_k) {\mathcal G}_{jk}(s) \right ]\left ( 1 - \frac{\rho_j}{|\y|} \right ) +\frac{\overline{\chi}_j(s)}{|\y|}  \widetilde{\Psi}(1/\rho_j)+\frac{4\pi D}{|\y| }G_{j0}(s) \widetilde{\psi}(1/\rho_i)
{\mathcal G}_{jj}(s)\ \nonumber \\
&\quad +\frac{4\pi D}{|\y|} \left [ \sum_{k\neq j}G_{k0}(s) \left \{ \frac{\widetilde{\Psi}(1/\rho_k)-\widetilde{\Psi}(1/\rho_j)}
{\rho_j^{-1}-\rho_k^{-1}}\right \} {\mathcal G}_{jk}(s) \right ] +\mbox{ first-order spherical harmonics}.
\label{QPsi1}
\end{align}
\end{widetext}
We have introduced the stopping local time density 
\begin{equation}
\label{psi}
\psi(\ell)=-\frac{d\Psi(\ell)}{d\ell}, \quad \widetilde{\psi}(q)=1-q\widetilde{\Psi}(q).
\end{equation}
It can be checked that (\ref{QPsi0}) and (\ref{QPsi1}) recover Eq. (\ref{inner}), on setting $\Psi(\ell)=\e^{-\gamma\ell}$ and $\widetilde{\Psi}(q)=(q+\gamma)^{-1}$. 

\begin{figure*}[t!]
\centering
\includegraphics[width=18cm]{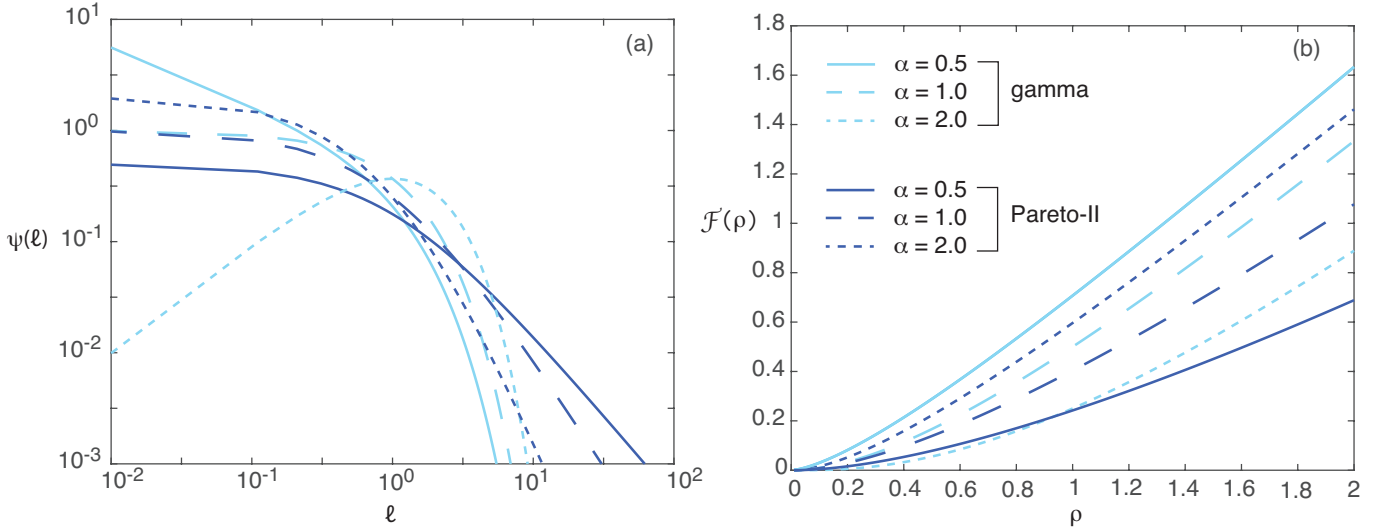} 
\caption{(a) Plots of the probability density $\psi(\ell)$ as a function of the stopping local time for the gamma and Pareto-II models. (b) Corresponding plots of the renormalized target radius ${\mathcal F}(\rho)$ as a function of the physical radius $\rho$. We also set $\gamma=\kappa_0/D=1$.}
\label{fig2}
\end{figure*}

\subsection{The generalized target flux}

Multiplying both sides of the boundary condition (\ref{prop}c) by $\Psi(\ell)$ and integrating by parts with respect to $\ell$ shows that
\begin{eqnarray}
-D\nabla p(\x,t|\x_0) \cdot \n_j=D\int_0^{\infty}\psi(\ell) P(\x,\ell,t|\x_0)d\ell \nonumber \\
\end{eqnarray}
for $ \x\in \partial \calU_j$. We have used Eq. (\ref{prop}d) and the identity $\Psi(0)=1$. Laplace transforming, introducing stretched coordinates and integrating with respect to points on the boundary $\partial \calU_j$ gives the flux into the $j$th target:
\begin{equation}
\label{Jg}
\widetilde{J}_j(\x_0,s)= D\epsilon \rho_j^2\int_0^{\infty}\psi(\ellh)\left [ \int_{\partial \calU_j}\V(\y,\ell,s|\x_0)d\sigma \right ]d\ellh.
\end{equation}
Substituting the asymptotic expansion (\ref{Vexp}) of the propagator then gives
\begin{align}
\label{JLTPsi}
&\widetilde{J}_j(\x_0,s)
\sim 4\pi \epsilon D\bigg [{\mathcal F}(\rho_j)G_{j0}(s) \\
&\quad -4\pi \epsilon D[{\mathcal F}(\rho_i)-\rho_j\widetilde{\psi}'(1/\rho_j)]G_{j0}(s){\mathcal G}_{jj}(s)\nonumber \\
&-4\pi \epsilon D\sum_{k\neq j}G_{k0}(s)\left \{ \frac{\rho_k^2{\mathcal F}(\rho_j)-\rho_j^2{\mathcal F}(\rho_k)}{\rho_k-\rho_j}\right \}{\mathcal G}_{jk}(s) 
\bigg ] \nonumber \\
&\quad +O(\epsilon^3),\nonumber\end{align}
where
\begin{equation}
{\mathcal F}(\rho)=\rho-\widetilde{\Psi}(1/\rho).
\end{equation}
We have used Eq. (\ref{psi}), which  implies that
\begin{equation}
\rho_j\widetilde{\psi}(1/\rho_j)=\rho_j\left (1-\frac{1}{\rho_j}\widetilde{\Psi}(1/\rho_j)\right )={\mathcal F}(\rho_j).
\end{equation}
Hence, the leading order terms involve an effective renormalization of the target size.

Taking the limit $s\rightarrow 0$ in Eq. (\ref{JLTPsi}) yields a corresponding asymptotic expansion of the splitting probabilities:
\begin{align}
 &\pi_j(\x_0)=\lim_{s\rightarrow 0}\widetilde{J}_j(\x_0,s) \\
 &= \epsilon  {\mathcal F}(\rho_j)\bigg [\frac{1}{|\x_j-\x_0|}\nonumber \\
 &\quad - \epsilon \sum_{k\neq j} \frac{1}{|\x_k-\x_0| }\left \{ \frac{\rho_k^2{\mathcal F}(\rho_j)-\rho_j^2{\mathcal F}(\rho_k)}{\rho_k-\rho_j}\right \}\frac{1}{|\x_k-\x_j|}\bigg]   \nonumber\\
 &\quad  +O(\epsilon^3).\nonumber
\label{splitpsi}
\end{align}
 For the sake of illustration, we list a few possible surface reaction models in terms of the probability density $\psi(\ell)$
and the equivalent encounter-dependent reactivity $\kappa(\ell)$ defined in Eq. (\ref{kaell}). See Table 1 of Ref. \cite{Grebenkov20} for a more comprehensive list. In each case we take $\gamma=\kappa_0/D$ where $\kappa_0$ is some reference reactivity.
\medskip

\begin{figure}[b!]
\centering
\includegraphics[width=8.5cm]{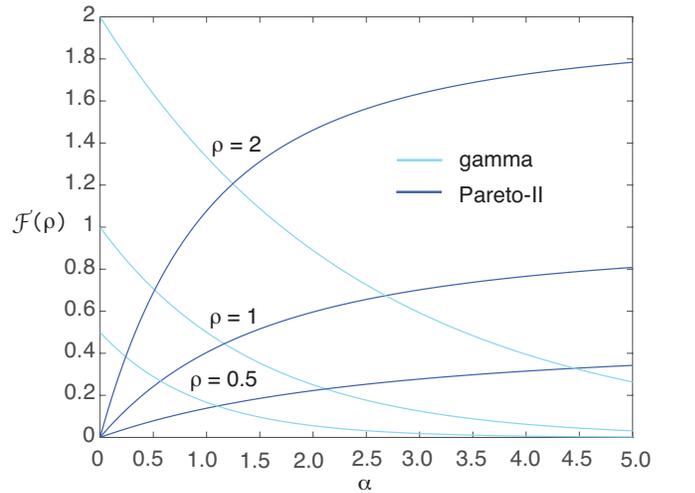} 
\caption{Plots of ${\mathcal F}(\rho)$ for $\rho=0.5,1,2$ as a function of the coefficient $\alpha$ for the gamma and Pareto-II models.}
\label{fig3}
\end{figure}

\begin{figure}[t!]
\centering
\includegraphics[width=6cm]{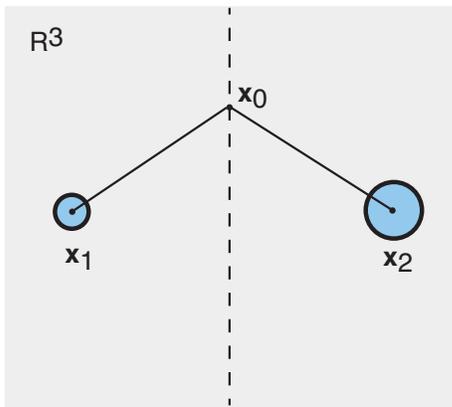} 
\caption{Two spherical targets of radii $\rho_1$ and $\rho_2$. For simplicity, the initial position is taken to be equidistant from the centers of the two targets.}
\label{fig4}
\end{figure}

\begin{figure*}[t!]
\centering
\includegraphics[width=18cm]{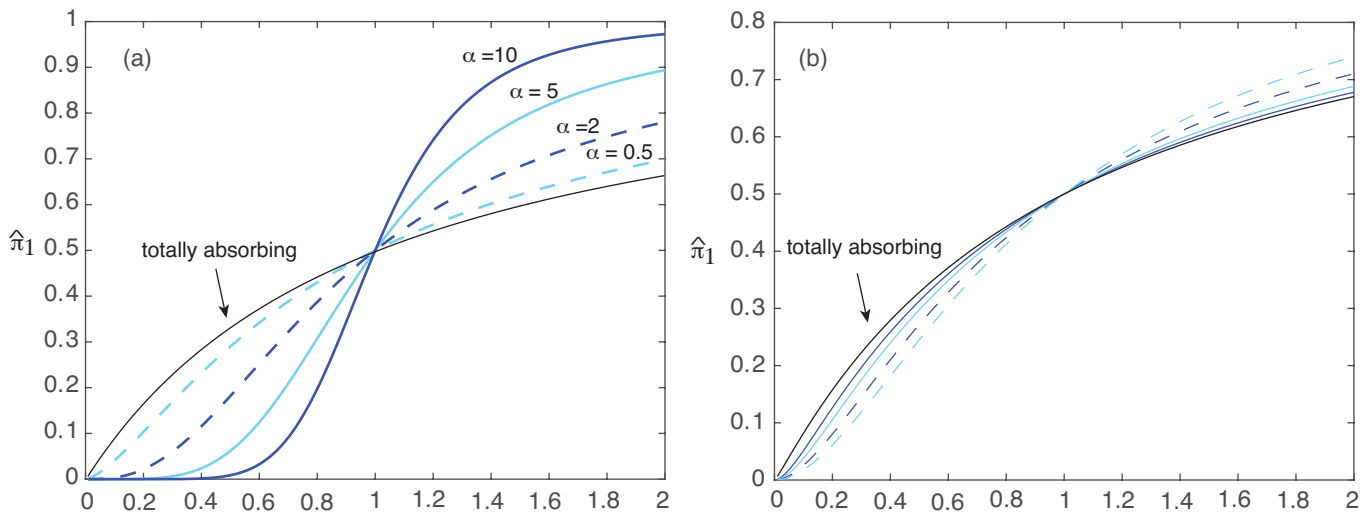} 
\caption{Two spherical targets with rescaled radii $\rho_1$ and $\rho_2$. Plot of leading-order contribution to the normalized splitting probability of the first target, $\widehat{\pi}_1\sim {\mathcal F}(\rho_1)/({\mathcal F}(\rho_1)+{\mathcal F}(\rho_2))$, as a function of $\rho_1$ for  $\rho_2=1$ and $\gamma=1$. (a) $\Psi$ given by the gamma distribution. (b) $\Psi$ given by the Pareto-II distribution. We also set $\gamma=\kappa_0/D=1$. Also shown is the normalized splitting probability for totally absorbing targets ($\kappa_0\rightarrow \infty$).}
\label{fig5}
\end{figure*}

\noindent {\em (a) Exponential distribution.}
\begin{equation}
\psi(\ell)=\gamma \e^{-\gamma \ell},\quad \widetilde{\psi}(q)=\frac{\gamma}{\gamma+q},\quad \kappa(\ell) =\kappa_0.
\end{equation}

\noindent  {\em (b) Gamma distribution.}
\begin{subequations}
\begin{equation}
\psi(\ell)=\frac{\gamma(\gamma \ell)^{\alpha-1}\e^{-\gamma \ell}}{\Gamma(\alpha)},\quad \widetilde{\psi}(q)=\left (\frac{\gamma}{\gamma+q}\right )^{\alpha},\
\end{equation}
and
\begin{equation}\kappa(\ell)=\kappa_0 \frac{ (\gamma \ell)^{\alpha-1}\e^{-\gamma \ell}}{\Gamma(\alpha,\gamma \ell)},\end{equation}
where $\Gamma(\alpha)$ is the gamma function and $\Gamma(\alpha,z)$ is the upper incomplete gamma function.
\end{subequations}
\medskip

\noindent  {\em (c) Pareto-II (Lomax) distribution.}
\begin{subequations}
\begin{equation}
\psi(\ell)=\frac{\gamma \alpha}{(1+\gamma \ell)^{1+\alpha}},\quad \widetilde{\psi}(q)=\alpha\left (\frac{q}{\gamma}\right )^{\alpha}\e^{q/\gamma}\Gamma(-\alpha,q/\gamma),
\end{equation}
and
\begin{equation}\kappa(\ell)=\kappa_0 \frac{\alpha}{1+\gamma \ell}.
\end{equation}
\end{subequations}

In Fig. \ref{fig2}(a) we plot the probability density $\psi(\ell)$ as a function of the stopping local time $\ell$ for the gamma and Pareto-II models and the particular coefficients $\alpha=0.5,1,2$. We also set $\gamma=1$. (The gamma density for $\alpha=1$ gives the exponential model). In Fig. \ref{fig2}(b) we show the corresponding plots of the renormalized target radius function ${\mathcal F}(\rho)=\rho-\widetilde{\Psi}(\rho)$. In all cases, ${\mathcal F}(\rho)$ is a nonlinear, monotonically increasing function of $\rho$. Moreover, ${\mathcal F}(\rho)$ is sensitive to the value of the $\alpha$-coefficient that parameterizes each of the two probability distributions. That is, ${\mathcal F}(\rho)$ is a decreasing (increasing) function of $\alpha$ for fixed $\rho$ in the case of the gamma (Pareto-II) model. Having determined the renormalized radius ${\mathcal F}(\rho)$, we can now explore how the choice of surface reaction model modifies the leading-order contributions to the splitting probabilities for more than one target.

For the sake of illustration, consider two spherical targets of rescaled radii $\rho_1$ and $\rho_2$ such that $\rho_2$ is fixed at unity, see Fig. \ref{fig4}. Assuming that the target centers are equidistant from the starting position $\x_0$, that is $|\x_1-\x_0|=|\x_2-\x_0|$, the normalized splitting probabilities $\widehat{\pi}_j=\pi_j/(\pi_1+\pi_2)$ are 
\begin{equation}
\widehat{\pi}_j=\frac{{\mathcal F}(\rho_j)}{{\mathcal F}(\rho_1)+{\mathcal F}(\rho_2)}.
\end{equation}
 In Fig. \ref{fig5} we plot the leading-order contribution to the normalized splitting probability $\widehat{\pi}_1$ of the first target as a function of the target radius $\rho_1$ for the gamma and Pareto-II models. As expected, $\pi_1=0.5$ when $\rho_1=\rho_2=1$. In the case of the gamma model, $\widehat{\pi}_1$ is a sigmoid-like function of $\rho_1$ whose steepness increases significantly with the $\alpha$-coefficient. That is, for large $\alpha$, small changes in $\rho_1$ leads to large changes in the renormalized radius, and thus $\widehat{\pi}_1$. The latter effect is much weaker in the case of the Pareto-II model. 

\setcounter{equation}{0}\section{Discussion}  
 In this paper we analyzed the 3D narrow capture problem for small spherical targets with partially reactive boundary surfaces. We proceeded by combining matched asymptotic analysis with an encounter-based formulation of diffusion-mediated surface reactions. In particular, we derived an asymptotic expansion of the joint probability density (propagator) for the position and boundary local time of reflected Brownian motion. The effects of surface reactions were then incorporated via an appropriate stopping condition for the boundary local time. We illustrated the theory by investigating how surface reactions affected the splitting probabilities. We showed that to leading order there is an effective renormalization of the target radius of the form $\rho\rightarrow \rho-\widetilde{\Psi}(1/\rho)$, where $\widetilde{\Psi}$ is the Laplace transform of the stopping local time distribution.
 
In order to facilitate the analysis, we made a number of simplifying assumptions. First, the region $\Omega$ containing the targets was taken to be unbounded, that is, $\Omega=\R^3$. The analysis of the target fluxes in the small-$s$ limit is considerably more involved when $\Omega$ is bounded. Suppose, in particular, that the exterior boundary $\partial \Omega$ is totally reflecting. The corresponding Neumann Green's function of the modified Helmholtz equation in $\Omega$ then has a singularity of the form $G(\x,s|\x_0)\sim 1/s$. In the case of totally absorbing targets, the resulting singularities in the asymptotic expansion of the Laplace transformed fluxes can be eliminated by considering a triple expansion in $\epsilon$, $s$ and $\Gamma \propto \epsilon /s$ \cite{Bressloff21b}. Performing partial summations over infinite power series in $\Gamma$ leads to multiplicative factors of the form $\Gamma^n/(1+\Gamma)^n $. Since $\Gamma^n/(1+\Gamma)^n \rightarrow 1$ as $s\rightarrow 0$, the singularities in $s$ are removed. However, extending this analysis to partially reflecting targets is non-trivial. 

Another major difference between unbounded and bounded domains $\Omega$ is that the splitting probabilities are $O(1)$ rather than $O(\epsilon)$ and $\sum_{j=1}^N\pi_k=1$. Moreover, one can now construct conditional mean first passage times (MFPTs); these are infinite when $\Omega=\R^3$. The FPT $\calT_k$ to be captured by the $k$-th target is 
\begin{equation}
\calT_k(\x_0)=\inf\{t>0; \X(t)\in \partial\calU_k|\X(0)=\x_0\},
\end{equation}
with $\calT_k=\infty$ if the particle is captured by another target. Introducing the set of events $\Omega_k=\{\calT_k<\infty\}$, the conditional FPT densities are defined according to
\begin{align*}
f_k(\x_0,t)dt&=\P[t<\calT_k<t+dt|\calT_k<\infty,\X(0)=\x_0].
\end{align*}
One finds that
\begin{equation}
\label{fk}
f_k(\x_0,t)=\frac{J_k(\x_0,t)}{\pi_k(\x_0)}.
\end{equation} 
Moreover, the Laplace transform of $f_k(\x_0,t)$ is the generator of the moments of the conditional FPT density:
\begin{align}
\label{fkLT}
\E[\e^{-s\calT_k}|1_{\Omega_k}]=\widetilde{f}_k(\x_0,s)=\frac{\widetilde{J}_k(\x_0,s)}{\widetilde{J}_k(\x_0,0)},
\end{align}
and
\begin{align}
T_k^{(n)}&=\E[\calT_k^n|1_{\Omega_k}]=\left . \left (-\frac{d}{ds}\right )^n\E[\e^{-s\calT_k}|1_{\Omega_k}]\right |_{s=0}\nonumber \\
&=\left . \left (-\frac{d}{ds}\right )^n\widetilde{f}_k(\x_0,s)\right |_{s=0}.
\end{align}
In particular, the conditional MFPT $T_k=T_k^{(1)}$ is  
\begin{align}
\label{mfpt}
\pi_k(\x_0)T_k(\x_0)  = \lim_{s\rightarrow 0} \left .\frac{d\widetilde{J}_k(\x_0,s)}{ds}\right |_{s=0}. 
\end{align}
As with the splitting probabilities, the calculation of $T_k(\x_0) $ requires taking the limit $s\rightarrow 0$ and hence dealing with the singular nature of the Green's function.

A second simplifying assumption was to consider spherically-shaped targets. However, as originally shown by Ward and Keller \cite{Ward93,Ward93a}, it is possible to generalize the asymptotic analysis of narrow capture problems to more general target shapes such as ellipsoids by applying classical results from electrostatics. In the case of totally absorbing targets one simply replaces the target length $\rho_j$ in the far-field behavior of the inner solution by the capacitance $C_j$ of an equivalent charged conductor with the shape $\calU_j$. In addition, using the divergence theorem, it can be shown that the flux into a target is completely determined by the far-field behavior. It would be interesting to determine the effective renormalization of the capacitances in the case of partially absorbing targets. A third simplification was to take the rule for surface reactions to be the same for each target, which meant that we only needed to keep track of a single boundary local time. If each target were to have a different probability distribution for the stopping local time, then it would be necessary to introduce multiple local times $\ell_j$, $j=1,\ldots,N$ \cite{Grebenkov20a}. The associated propagator would then be $P=P(\x,\ell_1,\ldots\ell_N,t|\x_0)$ such that the marginal probability density becomes
\begin{align*}
p(\x,t|\x_0)&=\int_0^{\infty}d\ell_1 \Psi_1(\ell_1) \ldots \int_0^{\infty}d\ell_N\Psi_N(\ell_N)\\
&\quad \times P(\x,\ell_1,\ldots\ell_N,t|\x_0).
\end{align*}

Finally, note that a complementary approach to dealing with partially reactive surfaces arises within the context of multi-scale computational models of reaction-diffusion (RD) systems. A major challenge in simulating intracellular processes is how to efficiently couple stochastic chemical reactions involving low molecular numbers with diffusion in complex environments. One approach is to consider a spatial extension of the Gillespie algorithm for well-mixed chemical reactions \cite{Gillespie77,Gillespie01} using a mesoscopic compartment-based method, although there are subtle issues with regards choosing the appropriate compartment size  \cite{Turner04,Isaacson06,Isaacson09,Hu13}. Alternatively, one can combine a coarse-grained deterministic RD model in the bulk of the domain with individual particle-based Brownian dynamics in certain restricted regions \cite{Andrews04,Erban07,Erban09,Franz13}; in this case considerable care must be taken in the choice of boundary conditions at the interface between the two domains. This is somewhat analogous to having to deal with boundary local times in partially reflecting Brownian motion.

\renewcommand{\theequation}{A.\arabic{equation}}

\setcounter{equation}{0}

 \section*{Appendix A. Derivation of the propagator BVP using a Feynman-Kac formula}
 
 Another way to define the propagator $P(\x,\ell,t|\x_0)$ introduced in Sect. III is in terms of the expectation of a Dirac delta function with respect to the distribution of
 paths between $(\x_0,0)$ to $(\x,t)$:
 \begin{align}
 \label{A1}
& P(\x,\ell,t|\x_0)=\bigg \langle \delta\left (\ell -D{\mathcal T}(\partial\calU,t)\right )\bigg \rangle_{\X_0=\x_0}^{\X_t=\x} ,
 \end{align}
 where
 \begin{equation}
{\mathcal T}(\partial\calU,t)= \int_0^t\int_{\partial \calU} \delta(\X_{\tau}-\x)d\x d\tau.
 \end{equation}
 That is, the joint probability density is obtained by summing over all paths whose accumulative boundary local time is equal to $\ell$.
 Using a Fourier representation of the Dirac delta function, Eq. (\ref{A1}) can be rewritten as
 \begin{align}
 P(\x,\ell,t|\x_0)=\int_{-\infty}^{\infty} \e^{i\omega \ell}{\mathcal G}(\x,\omega,t|\x_0)\frac{d\omega}{2\pi},
 \end{align}
 where $ P(\x,\ell,t|\x_0)=0$ for $\ell <0$ and
 \begin{align}
 {\mathcal G}(\x,\omega,t|\x_0)=\bigg\langle \exp \left ( -i\omega D{\mathcal T}(\partial\calU,t)\right )\bigg \rangle_{\X_0=\x_0}^{\X_t=\x}.
 \end{align}
 We now note that ${\mathcal G}$ is the characteristic functional of the Brownian local time, which can be evaluated using a path-integral representation of the stochastic process. The latter can then be used to derive the following Feynman-Kac equation \cite{Kac49,Majumdar05}:
\begin{align}
\label{calGa}
\frac{\partial \calG(\x,\omega,t|\x_0)}{\partial t}&=D\nabla^2 \calG(\x,\omega,t|\x_0)\\
&\quad -i\omega D \int_{\partial \calU} \calG(\x',\omega,t|\x_0) \delta(\x-\x')d\x'.\nonumber
\end{align}

Multiplying Eq. (\ref{calGa}) by $\e^{\i\omega\ell}$, integrating with respect to $\omega$ and using the identity
\[\frac{\partial }{\partial \ell} P(\x,\ell,t|\x_0)\Theta(\ell) =\int_{-\infty}^{\infty} i\omega D \e^{i\omega \ell}{\mathcal G}(\x,\omega,t|\x_0)\frac{d\omega}{2\pi},\]
with $\Theta(\ell)$ the Heaviside function, we obtain the result
\begin{align}
\label{calP}
\frac{\partial P(\x,\ell,t|\x_0)}{\partial t}&=D\nabla^2 P(\x,\ell,t|\x_0)\\
&\quad - D\int_{\partial \calU} \frac{\partial P}{\partial \ell}(\x',\ell,t|\x_0) \delta(\x-\x')d\x'\nonumber\\
&\quad - D\delta(\ell)\int_{\partial \calU}P(\x',0,t|\x_0) \delta(\x-\x')d\x'\nonumber .
\end{align}
This is equivalent to the BVP
\begin{align*}
&\frac{\partial P(\x,\ell,t|\x_0)}{\partial t}=D\nabla^2 P(\x,\ell,t|\x_0),\ \x \in \R^3\backslash \calU \\
&-D\nabla P(\x,\ell,t|\x_0) \cdot \n= D P(\x,\ell=0,t|\x_0) \ \delta(\ell) \nonumber \\
& \quad +D\frac{\partial}{\partial \ell} P(\x,\ell,t|\x_0),\  \x\in \partial \calU,\nonumber
\end{align*}
which reduces to Eq. (\ref{prop0}) on setting $P(\x,\ell=0,t|\x_0)=-\nabla p_{\infty}(\x,t|\x_0)\cdot \n $ for $\x\in \partial \calU$. The latter equality can be understood by noting that a constant reactivity is equivalent to a Robin boundary condition. Thus
\begin{align}
\nabla p(\x,t|\x_0)\cdot \n&=-\kappa_0 p(\x,t|\x_0)\nonumber \\
&=-\kappa_0 \int_0^{\infty}\e^{-\kappa_0 \ell}P(\x,\ell,t|\x_0)d\ell.
\end{align}
The result follows from taking the limit $\kappa_0 \rightarrow \infty$ on both sides and noting that $\lim_{\kappa_0 \rightarrow \infty}\kappa_0 \e^{-\kappa_0 \ell}$ is the Dirac delta function on the positive half-line.


\begin{thebibliography}{9}

\bibitem{Ward93} {M. J. Ward  and J. B. Keller,} {Strong localized perturbations of eigenvalue problems.} {SIAM J Appl Math} {\bf 53} 770-798 (1993).

\bibitem{Ward93a} {M. J. Ward , W. D. Henshaw and J. B. Keller,} {Summing logarithmic expansions for singularly perturbed eigenvalue problems.} SIAM J. Appl. Math. {\bf 53 } 799-828 (1993).

\bibitem{Ward00} {M. J. Ward,}  {Diffusion and bifurcation problems in singularly perturbed domains.} 
Natural Resource Modeling {\bf 13} (2000).


\bibitem{Straube07} { R. Straube, M. J. Ward and M. Falcke,} {Reaction rate of small diffusing molecules on a cylindrical membrane.} {J. Stat. Phys.} {\bf 129} 377-405 (2007).



\bibitem{Schuss07} {Z. Schuss, A. Singer and D.Holcman,} {The narrow escape
problem for diffusion in cellular microdomains.} {Proc. Natl.
Acad. Sci. (U.S.A.)} {\bf 104} 16098 (2007).

\bibitem{Bressloff08} {P. C. Bressloff, B. A. Earnshaw and M. J. Ward,} {Diffusion of protein receptors on a cylindrical dendritic membrane with partially absorbing targets.} {SIAM J. Appl. Math.} {\bf 68} 1223-1246 (2008).



  \bibitem{Benichou08}
{ O. Benichou and R. Voituriez} {Narrow escape time problem: Time needed for a
  particle to exit a confining domain through a small window.}
\newblock {Phys. Rev. Lett.} {\bf 100} 168105 (2008).





\bibitem{Coombs09} {D. Coombs, R. Straube and M. J. Ward,} {Diffusion on a sphere with localized targets: Mean first passage time, eigenvalue asymptotics, and Fekete points.} {SIAM J. Appl. Math.} {\bf 70} 302-332 (2009).

 \bibitem{Pillay10}
{S. Pillay, M. J. Ward, A. Peirce, T.  Kolokolnikov,} {An asymptotic analysis of the mean first passage time for narrow escape problems: {P}art {I}:
  Two-dimensional domains.}
\newblock {SIAM Multiscale Model. Sim.} {\bf 8} 803-835 (2010).

\bibitem{Reingruber10} {J. Reingruber and D. Holman} {Narrow escape for a stochastically gated Brownian ligand.} J. Phys. Cond. Matter {\bf 22} 065103 (2010).




\bibitem{Cheviakov10} {A. F. Cheviakov, M. J. Ward and R. Straube} {An asymptotic analysis of the mean first passage time for narrow escape problems: {P}art {II}: {T}he sphere.} {SIAM J. Multiscal Mod. Sim.} {\bf 8} 836-870 (2010).


\bibitem{Cheviakov11} {A. F. Cheviakov and M. J. Ward} {Optimizing the principal eigenvalue of the Laplacian in a sphere with interior targets.} {Math.
Comp. Modeling} {\bf 53} 042118 (2011).


\bibitem{Chevalier11} {C. Chevalier, O. Benichou, B. Meyer and R. Voituriez} {First-passage quantities of Brownian motion in a bounded domain with multiple targets: a unified approach.} {J. Phys. A} {\bf 44} 025002 (2011).

\bibitem{Holcman14a} {D. Holcman and Z. Schuss,} {The narrow escape problem} {SIAM Rev.} {\bf 56} 213 (2014)

\bibitem{Ward15} {V. Kurella, J. C. Tzou, D. Coombs and M. J. Ward,} 
{Asymptotic analysis of first passage time problems
inspired by ecology.} {Bull Math Biol.} {\bf 77} 83-125 (2015).

\bibitem{Coombs15} { M. I. Delgado, M. J. Ward  and D. Coombs,} {Conditional mean first passage times to small targets in a 3-D domain with a sticky boundary: Applications to T cell searching behavior in lymph nodes.} {Multiscale Model. Simul.} {\bf 13} 1224-1258 (2015)..


\bibitem{Bressloff15} {P. C. Bressloff and S. D. Lawley,} {Stochastically-gated diffusion-limited reactions for a small target in a bounded domain.} {Phys. Rev. E} {\bf 92} 062117 (2015).

\bibitem{Bressloff15a} {P. C. Bressloff and S. D. Lawley,} {Escape from subcellular domains with randomly switching boundaries.} {Multiscale Model. Simul.} {\bf 13} 1420-1445 (2015).

\bibitem{Lindsay15} {A. E. Lindsay , T.  Kolokolnikov and J. C. Tzou,} {Narrow escape problem with a mixed target and the effect of orientation.} {Phys. Rev. E} {\bf 91} 032111 (2015).
   
 \bibitem{Lindsay16} {A. E. Lindsay, R. T. Spoonmore and J. C. Tzou,} {Hybrid asymptotic-numerical approach for estimating first passage time densities of the two-dimensional narrow capture problem.} {Phys. Rev. E} {\bf 94} 042418 (2016).
 
 \bibitem{Lindsay17} {A. E. Lindsay, A. J. Bernoff and M. J. Ward,} {First passage statistics for the capture of a Brownian particle by a structured spherical target with multiple surface targets.} {Multiscale Model. Simul.} {\bf 15} 74-109 (2017).


\bibitem{Grebenkov17} {D. S. Grebenkov and G. Oshanin,} {Diffusive escape through a narrow
opening: new insights into a classic problem.} Phys. Chem. Chem. Phys.
{\bf 19} 2723-2739 (2017).


\bibitem{Bressloff21a} {P. C. Bressloff,} {Asymptotic analysis of extended two-dimensional narrow capture problems}. Proc Roy. Soc. A {\bf 477} 20200771 (2021). 

\bibitem{Bressloff21b} {P. C. Bressloff,} {Asymptotic analysis of target fluxes in the three-dimensional narrow capture problem}. Multiscale Model. Simul. {\bf 19 } 612-632 (2021).  

\bibitem{Bressloff22a} {P. C. Bressloff,} {Accumulation time of diffusion in a singularly perturbed domain}. Preprint (2022).

\bibitem{Holcman15} {D. Holcman and Z. Schuss,}  {Stochastic Narrow Escape in Molecular and
Cellular Biology} (Springer, New York, 2015).



\bibitem{Bressloff22} {P. C. Bressloff,} {Stochastic Processes in Cell Biology: Vols. I and II} (Springer 2022).




\bibitem{Rice85}
{S. A. Rice,} {Diffusion-limited reactions}.
\newblock (Elsevier, Amsterdam 1985).



\bibitem{Collins49}
{F. C. Collins and G. E. Kimball,} {Diffusion-controlled reaction rates.}
\newblock J. Colloid Sci. {\bf 4} 425-439 (1949).


\bibitem{Lawley15} {S. D. Lawley and J. P. Keener,} {A new derivation of Robin boundary conditions through homogenization of a stochastically switching boundary}. SIAM Journal on Applied Dynamical Systems {\bf 14} 1845-1867 (2015).


\bibitem{Grebenkov19a} {D. S. Grebenkov} {Imperfect Diffusion-Controlled Reactions.}
in Chemical Kinetics: Beyond the Textbook, Eds.
K. Lindenberg, R. Metzler, and G. Oshanin (World Scientific,
2019).

\bibitem{Levy39} {P. L\`evy} {Sur certaines processus stochastiques homogenes.} Compos. Math. {\bf 7} 283 (1939).

 
 \bibitem{McKean75} {H. P. McKean} {Brownian local time.} Adv. Math. {\bf 15} 91-111 (1975).
 
 \bibitem{Freidlin85} {M. Freidlin.} {Functional Integration and Partial Differential Equations}
Annals of Mathematics Studies (Princeton University Press, Princeton,
New Jersey, 1985).

\bibitem{Papanicolaou90} {V. G. Papanicolaou} {The probabilistic solution of the third boundary
value problem for second order elliptic equations} Probab. Th. Rel. Fields
{\bf 87} 27-77 (1990).
 
 
 \bibitem{Milshtein95} {G. N. Milshtein} {The solving of boundary value problems by numerical
integration of stochastic equations.} Math. Comp. Sim. {\bf 38} 77-85 (1995).





\bibitem{Grebenkov03} {D. S. Grebenkov, M. Filoche, and B. Sapoval.} {Spectral Properties of the
Brownian Self-Transport Operator.} Eur. Phys. J. B {\bf 36}  221-231 (2003).

\bibitem{Grebenkov06} {D. S. Grebenkov.} {Partially reflected Brownian motion: A stochastic
approach to transport phenomena.} in Focus on Probability Theory,
Ed. L. R. Velle, pp. 135-169 (Hauppauge: Nova Science Publishers,
2006).

\bibitem{Grebenkov07} {D. S. Grebenkov.} {Residence times and other functionals
of reflected Brownian motion.} Phys. Rev. E {\bf } 041139 (2007).


\bibitem{Singer08} {A. Singer, Z. Schuss, A. Osipov, and D. Holcman.} {Partially reflected
diffusion.} SIAM J. Appl. Math. {\bf 68} 844-868 (2008).

\bibitem{Grebenkov19b} {D. S. Grebenkov} {Spectral theory of imperfect diffusion-controlled
reactions on heterogeneous catalytic surfaces}
J. Chem. Phys. {\bf 151} 104108 (2019).

\bibitem{Grebenkov20} {D. S. Grebenkov} {Paradigm shift in diffusion-mediated surface phenomena.} Phys. Rev. Lett. {\bf 125}  078102 (2020).


\bibitem{Grebenkov21} {D. S. Grebenkov} {An encounter-based approach for restricted diffusion with a gradient drift.}  arXiv:2110.12181 (2021).

\bibitem{Bressloff22} P. C. Bressloff, Diffusion-mediated surface reactions, Brownian functionals and the Feynman-Kac formula. Preprint (2022).

\bibitem{Grebenkov20a} {D. S. Grebenkov}, Joint distribution of multiple boundary local times and related first-passage time problems with multiple targets. Journal of Statistical Mechanics: Theory and Experiment {\bf 10} 103205 (2020).

\bibitem{Gillespie77} D. T. Gillespie,  Exact stochastic simulation of coupled chemical reactions. J. Phys. Chem. {\bf 81} 2340-2361 (1977).

\bibitem{Gillespie01} D. T. Gillespie, Approximate accelerated stochastic simulation of chemically reacting systems. J. Chem. Phys. {\bf 115},1716-1733 (2001).

\bibitem{Turner04} T. E. Turner, S. Schnell and K. Burrage, Stochastic approaches for modelling in vivo reactions. Comp. Biol. Chem. {\bf 28} 165-178 (2004).

 
\bibitem{Isaacson06} S. A. Isaacson and C. Peskin, Incorporating diffusion in complex geometries into stochastic chemical
kinetics simulations, SIAM J. Sci. Comp. {\bf 28} 47-74 (2006).

\bibitem{Isaacson09} S. A. Isaacson, The reaction-diffusion master equation as an asymptotic approximation of diffusion
to a small target. SIAM J. Appl. Math. {\bf 7} 77-111 (2009).

\bibitem{Hu13} J. Hu, H.-W. Kang and H. G. Othmer, Stochastic analysis of reaction-diffusion processes. Bull. Math. Biol. {\bf 76} 854-894 (2014)


\bibitem{Andrews04} S. Andrews and D. Bray, Stochastic simulation of chemical reactions with spatial resolution
and single molecule detail. Phys. Biol. {\bf 1} 137-151 (2004).



\bibitem{Erban07} R. Erban and S. J. Chapman, Reactive boundary conditions for stochastic simulations of reaction-diffusion processes. Phys. Biol. {\bf 4} 16-28 (2007).

\bibitem{Erban09} R. Erban and S. J. Chapman, Stochastic modelling of reaction-diffusion processes: algorithms for bimolecular reactions. Phys. Biol. {\bf 6} 046001 (2009).

 \bibitem{Franz13} B. Franz, M. B. Flegg, S. J. Chapman and R. Erban, Mutiscale reaction-diffusion algorithms: {PDE}-assisted Brownian dynamics. SIAM J. Appl. Math. {\bf 73} 1224-1247 (2013).

\bibitem{Kac49} M. Kac, On distribution of certain Wiener functionals. Trans. Am. Math. Soc. {\bf 65}, 1-13 (1949).

\bibitem{Majumdar05}  S. N. Majumdar, Brownian functionals in physics and computer science. Curr. Sci. {\bf 89}, 2076 (2005).

 








\end{thebibliography}
\end{document}